\pdfoutput=1
\documentclass[11pt]{article}
\usepackage{subfig}
\usepackage{amssymb,amsmath,amsthm}
\usepackage{graphics,epsfig}
\usepackage{algorithmic}
\usepackage{algorithm}
\usepackage[margin=1 in]{geometry}

\usepackage{lipsum}
\usepackage{mathtools}
\usepackage{cuted}
\usepackage{hhline}
\usepackage{booktabs}
\usepackage{array}
\usepackage{placeins}
\usepackage[hidelinks]{hyperref}

\newtheorem{lem}{Lemma}[section]
\newtheorem{defs}[lem]{Definition}

\newtheorem{coro}[lem]{Corollary}
\newtheorem{thm}[lem]{Theorem}

\newtheorem{prop}[lem]{Proposition}
\newtheorem{fact}[lem]{Fact}
\theoremstyle{remark}
\newtheorem{rem}[lem]{Remark}
\newtheorem{exa}[lem]{Example}

\makeatletter
\def\bstctlcite{\@ifnextchar[{\@bstctlcite}{\@bstctlcite[@auxout]}}
\def\@bstctlcite[#1]#2{\@bsphack
  \@for\@citeb:=#2\do{%
    \edef\@citeb{\expandafter\@firstofone\@citeb}%
    \if@filesw\immediate\write\csname #1\endcsname{\string\citation{\@citeb}}\fi}%
  \@esphack}
\makeatother

\newcommand{\Cmax}{C_{\max}}
\newcommand{\OPT}{\mathrm{OPT}}
\newcommand{\Aone}{\textup{(A1)}}
\newcommand{\Aonep}{\textup{(A1$'$)}}
\newcommand{\Aonec}{\textup{(A1$^{\circ}$)}}
\newcommand{\Atwo}{\textup{(A2)}}
\newcommand{\Athree}{\textup{(A3)}}
\newcommand{\busy}{\mathrm{busy}}
\newcommand{\SAR}{\textup{\textsc{Sar}}}
\newcommand{\LIST}{\textup{\textsc{List}}}
\newcommand{\Hin}{\textup{(H)}}
\newcommand{\Win}{\textup{(W)}}

\title{Speedup-Aware Rounding for Scheduling Malleable Tasks with Precedence Constraints}
\author{Chi-Yeh~Chen 
\\ Department of Computer Science and Information
Engineering, \\ National Cheng Kung University, \\
Taiwan, ROC. \\
chency@mail.csie.ncku.edu.tw.}

\begin{document}
\bstctlcite{IEEEexample:BSTcontrol}

\maketitle
\begin{abstract}
We study the scheduling of malleable tasks with precedence constraints on $m$ identical processors to minimize the makespan, assuming that the processing time of a task does not increase and its work does not decrease with the number of processors, and that the work is a convex function of the processing time. The standard two-phase method solves a linear program, rounds the fractional processing times to allotments, caps the allotments at a threshold $\mu$, and runs list scheduling. We propose \emph{speedup-aware rounding}, which rounds a task to its slower allotment exactly when this allotment is at most a factor $a$ slower than the fractional processing time. We determine the worst-case increase of the work under this rule exactly and show that no per-task rounding rule with time factor $a$ does better. With $\mu=\lceil\beta(a)m\rceil$ the algorithm achieves the approximation ratio $13(30+\sqrt{493})/225\approx3.0162$ for every $m$, which improves the previous bound $2+\sqrt2\approx3.4142$; a single task with linear speedup shows that its ratio is at least $\approx2.9558$. For concave speedup functions, which include the roofline, communication, Amdahl and power-law models, the ratio becomes $\approx2.9739$, which improves the ratio $\approx3.2919$ of Jansen and Zhang. We also revisit the strictly decreasing case of our earlier work (Chen and Chu, 2013): the reported ratio $2.9549$ relies on an elimination step that is not valid. In this case the algorithm coincides with that of Jansen and Zhang up to its parameters, and with parameters chosen by the corrected analysis it guarantees their ratio $\approx3.2919$. Finally, without convexity no per-task rounding improves the known time--work trade-off, and the analysis cannot go below $(3+\sqrt5)/2$.

\end{abstract}

\section{Introduction}\label{sec:introduction}
Many parallel applications are built from computational kernels that are parallel themselves and that depend on one another through data flow. A common abstraction is a directed acyclic graph (DAG) of \emph{malleable tasks} in the sense of Turek, Wolf and Yu~\cite{Turek92}: before a task starts, the scheduler decides how many processors it receives; the task then runs on these processors without interruption, and its processing time depends on their number. Part of the parallel computing literature calls the same model \emph{moldable}~\cite{Benoit22,Perotin24}; we keep the terminology of~\cite{Lepere02b,Jansen06,Jansen11,Chen13}. Given $n$ malleable tasks, precedence constraints among them and $m$ identical processors, we look for a feasible schedule of minimum makespan.

The problem is strongly NP-hard already for independent tasks on five processors~\cite{Du89}. With precedence constraints it contains the classical problem $P|prec|C_{\max}$, so it cannot be approximated within $4/3$ unless $\mathrm{P}=\mathrm{NP}$~\cite{Lenstra78}, nor within $2-\varepsilon$ under a variant of the Unique Games Conjecture~\cite{Svensson11}. The difficulty that is specific to malleable tasks lies in the choice of the allotments: more processors shorten the critical paths but usually increase the total work. Choosing the allotments is a discrete time--cost tradeoff problem~\cite{Skutella98}. The deadline version of that problem admits no constant-factor approximation under the Unique Games Conjecture~\cite{Svensson13}, and instances of bounded depth $d$ cannot be approximated better than $(d+2)/4$ under the same conjecture~\cite{Daboul23}. Polynomial-time algorithms for malleable tasks therefore aim at bicriteria guarantees: a constant factor on the length of every path and a constant factor on the total work.

\subsection{Previous work}
\textbf{Two-phase algorithms.} Turek et al.~\cite{Turek92} proposed to first fix an allotment and then schedule the resulting tasks with fixed processor counts. For precedence constraints, Lep\`{e}re, Trystram and Woeginger~\cite{Lepere02b} combined this idea with a cap on large allotments and with list scheduling~\cite{Graham66}, and obtained the ratio $3+\sqrt5\approx5.236$ under the \emph{monotonous penalty assumptions} of Blayo et al.~\cite{Blayo99}: the processing time does not increase and the work does not decrease when a task receives more processors. Jansen and Zhang~\cite{Jansen06} computed the allotment from a linear program (LP), following Skutella's approach to time--cost tradeoffs~\cite{Skutella98}; they rounded the fractional processing times with a threshold parameter $\rho$, capped the allotments at a parameter $\mu$ and ran list scheduling, which gives a ratio of about $4.7306$ for optimized $\rho$ and $\mu$. In a second paper~\cite{Jansen11}, Jansen and Zhang assumed that the speedup $p_j(1)/p_j(l)$ of every task is concave in the number $l$ of processors. They showed that the work is then non-decreasing in $l$ and convex in the processing time, and obtained the ratio $100/63+100(\sqrt{6469}+13)/5481\approx3.291919$. Their LP divides by the differences $p_j(l)-p_j(l+1)$, so it needs processing times that strictly decrease in~$l$, as observed in~\cite{Chen13}; for a concave speedup whose processing time stops decreasing at some allotment, this can be arranged by discarding the allotments beyond that point (Lemma~\ref{lem:useless}). For strictly decreasing processing times, a concave speedup is in fact equivalent to a non-decreasing work that is convex in the processing time (Proposition~\ref{prop:concave}).

\textbf{Our earlier work.} Chen and Chu~\cite{Chen13} kept the monotonous penalty assumptions and added only the convexity of the work as a function of the processing time. They reformulated the LP of~\cite{Jansen06} so that the LP work of a task is the piecewise linear interpolation of its actual work, and proved the ratio $2+\sqrt2\approx3.4142$. For processing times that strictly decrease in the number of processors they reported the ratio $2.9549$; Section~\ref{sec:cc13} shows that this bound relies on an elimination step that is not valid, and that in this case their algorithm coincides with the one of~\cite{Jansen11} up to its parameters. Chen~\cite{Chen18} returned to the monotonous penalty model without convexity. Observing that the virtual work in the LP of~\cite{Jansen06} can underestimate the work of the allotment that it represents, he proposed an iterative method that solves a sequence of linear programs, each built from the solution of the previous one. He reported the ratio $4.4841$ after two iterations, which improves on $4.7306$, and a ratio that approaches $2+\sqrt2\approx3.4142$ (written $3.4143$ in~\cite{Chen18}) when the number of iterations is large (more than one hundred).

\textbf{Subsequent developments.} To the best of our knowledge, no smaller approximation ratio has been reported since~\cite{Chen13,Chen18} for the offline problem on arbitrary DAGs under these models; later work has turned to related settings. For precedence constraints of bounded width, G\"{u}nther, K\"{o}nig and Megow~\cite{Gunther14} gave a fully polynomial-time approximation scheme, and showed that the problem is strongly NP-hard when the precedence constraints form a tree. Makarychev and Panigrahi~\cite{Makarychev14} allowed preemption and considered concave processing functions; they obtained a $(2+\varepsilon)$-approximation, which is best possible under complexity-theoretic assumptions. Marchal et al.~\cite{Marchal18} studied task graphs whose tasks may change their allotment during execution under a practical speedup model with two thresholds; they proved that minimizing the makespan is NP-hard and showed that two heuristics are $2$-approximations. Perotin, Sun and Raghavan~\cite{Perotin21} and Perotin et al.~\cite{Perotin24jpdc} extended list-based scheduling of moldable workflows to $d$ types of resources, with ratio $1.619d+2.545\sqrt d+1$. In the online setting, where a task becomes known when all its predecessors have completed, Benoit et al.~\cite{Benoit22} and Perotin and Sun~\cite{Perotin24} obtained constant competitive ratios for the roofline, communication, Amdahl and general speedup models; Perotin and Sun also proved lower bounds for every online list scheduling algorithm whose allotment of a task depends only on the task itself. Perotin, Verrecchia and Raghavan~\cite{Perotin26} designed a single online algorithm for arbitrary monotonic speedup functions whose competitive ratio is $O(\sqrt P)$ on $P$ processors, and proved a lower bound of the same order. For the time--cost tradeoff problem, Daboul, Held and Vygen~\cite{Daboul23} gave a $d/2$-approximation for instances of depth~$d$. The speedup models used in the online works above satisfy all assumptions of this paper once useless allotments are removed (Section~\ref{sec:models}), so our offline guarantees apply to them.

\subsection{Contributions}
The rounding rules of~\cite{Jansen06,Jansen11,Chen13} look at the position of the fractional processing time $x_j^*$ of a task inside the interval spanned by two consecutive allotments, and round up when $x_j^*$ lies above a fixed fraction $\rho$ of that interval. Such a rule does not look at how much the faster allotment actually gains. Under the monotonous penalty assumptions, one more processor shortens a task by a factor of at most two, and a large gain is possible only when the additional processor costs almost no extra work. Speedup-aware rounding uses this observation directly: for a parameter $a>1$, a task is rounded to its slower allotment exactly when that allotment is at most $a$ times slower than $x_j^*$. Our results are the following; Table~\ref{tab:summary} gives an overview.

\begin{itemize}
\item \textbf{Tight per-task analysis.} Speedup-aware rounding increases a processing time by at most the factor $a$ and the work by at most a factor $b(a)$, which we determine exactly; for $13/9\le a<2$ it equals $1+\big((1-\sqrt{a-1})/(1+\sqrt{a-1})\big)^2$ (Theorem~\ref{thm:closed}). No rounding rule that guarantees the time factor $a$ for every task has a smaller worst-case work factor (Proposition~\ref{prop:optimal}).
\item \textbf{A bound for every $m$.} For every per-task rounding with time factor $a$ and work factor $b$, the cap $\mu=\lceil\beta(a)m\rceil$ with $\beta(a)=2/(a+2+\sqrt{a^2+4})$ gives the approximation ratio $a+b\,h(a)$, where $h(a)=(2-a+\sqrt{a^2+4})/2$, for every number $m$ of processors (Theorem~\ref{thm:general}). The proof bounds the min--max program of~\cite{Jansen06,Chen13} by weak LP duality and needs no case analysis on~$m$.
\item \textbf{Convex work.} Under \Aone, \Atwo\ and \Athree\ (defined below), the choice $a=13/9$ gives the ratio $13(30+\sqrt{493})/225\approx3.016208$, the best value of $a+b(a)h(a)$ (Theorem~\ref{thm:opt-a}); the simpler choice $a=3/2$, $\mu=\lceil m/3\rceil$ gives $57/2-18\sqrt2\approx3.044156$. Both improve the bound $2+\sqrt2$ of~\cite{Chen13}. A single task with linear speedup shows that the ratio of the algorithm with $a=13/9$ is at least $(31+\sqrt{493})/18\approx2.955756$, so the analysis is within $0.061$ of tight (Proposition~\ref{prop:lower}).
\item \textbf{Concave speedup.} If the processing time of every task strictly decreases until it reaches its minimum and then stays constant, the ratio becomes $\approx2.973912$ (Theorem~\ref{thm:strict} and Proposition~\ref{prop:strict-opt}). Together with \Atwo\ and \Athree, this condition is equivalent to a concave speedup function (Proposition~\ref{prop:concave}). The result improves the ratio $\approx3.2919$ of~\cite{Jansen11}; this includes concave speedup functions with a final plateau, such as the roofline model, to which~\cite{Jansen11} applies after the allotments beyond the plateau are discarded (Lemma~\ref{lem:useless}). Section~\ref{sec:models} shows that the roofline, communication, Amdahl, power-law and general speedup models belong to this class.
\item \textbf{The strictly decreasing case of~\cite{Chen13} revisited.} We identify the step of~\cite{Chen13} that does not hold. For strictly decreasing processing times the algorithm of~\cite{Chen13} coincides with that of~\cite{Jansen11} up to the parameters $\rho$ and $\mu$ (Section~\ref{sec:cc13-alg}). With parameters chosen by the corrected analysis, which are asymptotically those of~\cite{Jansen11}, it achieves $100/63+100(\sqrt{6469}+13)/5481\approx3.291919$, the ratio of~\cite{Jansen11} (Theorem~\ref{thm:cc13}). The guarantee extends to processing times that stay constant after reaching their minimum: the allotments beyond the minimum are never needed, and removing them leaves processing times that strictly decrease (Section~\ref{sec:cc13-alg}).
\item \textbf{Limits and lower bounds.} Without convexity, no per-task rounding of a relaxation whose work bounds are the convex hulls of the individual work functions achieves a better time--work trade-off than the one of~\cite{Jansen06} (Proposition~\ref{prop:a12}). Within the analysis framework, no rounding rule leads below $(3+\sqrt5)/2\approx2.618$ (Proposition~\ref{prop:frame}). The integrality gap of $(\mathrm{P})$ is at least $2$, and the cap alone forces the ratio $1/\beta(a)$; in particular the ratio $2+\sqrt2$ of the general rule of~\cite{Chen13} is tight (Proposition~\ref{prop:lower}).
\end{itemize}

\begin{table}[t]
\caption{Approximation ratios for every number $m$ of processors. \Aonec\ means that each processing time strictly decreases until it reaches its minimum (Section~\ref{sec:model}); together with \Atwo\ and \Athree\ it is equivalent to a concave speedup (Proposition~\ref{prop:concave}). The ratio of speedup-aware rounding is at least $\approx2.9558$ in the first row and $\approx2.8310$ in the second and third rows (Proposition~\ref{prop:lower}).}
\label{tab:summary}
\vspace{2mm}
\centering
\small
\begin{tabular}{>{\raggedright\arraybackslash}p{4.3cm}>{\raggedright\arraybackslash}p{4.9cm}>{\raggedright\arraybackslash}p{5.6cm}}
\toprule
Assumptions & Previous ratio & This paper\\
\midrule
\Aone, \Atwo, \Athree & $2+\sqrt2\approx3.4142$~\cite{Chen13} & $13(30+\sqrt{493})/225\approx3.0162$ (Theorem~\ref{thm:opt-a})\\
\Aonep, \Atwo, \Athree\ (strictly decreasing, concave speedup) & $\approx3.2919$~\cite{Jansen11} & $\approx2.9739$ (Theorem~\ref{thm:strict}, Proposition~\ref{prop:strict-opt})\\
\Aonec, \Atwo, \Athree\ (concave speedup, possibly with a final plateau) & $\approx3.2919$~\cite{Jansen11} after discarding useless allotments (Lemma~\ref{lem:useless}) & $\approx2.9739$ (Theorem~\ref{thm:strict}, Proposition~\ref{prop:strict-opt})\\
\Aonec, \Atwo, \Athree; algorithm of~\cite{Chen13} & $2.9549$ reported in~\cite{Chen13} under \Aonep & $\approx3.2919$ with parameters chosen by the corrected analysis (Theorem~\ref{thm:cc13})\\
\Aone, \Atwo & $\approx4.7306$~\cite{Jansen06}; $4.4841$ after two iterations, as reported in~\cite{Chen18} & limit of per-task rounding (Proposition~\ref{prop:a12})\\
\bottomrule
\end{tabular}
\end{table}

\textbf{Organization.} Section~\ref{sec:model} states the model, the assumptions and the LP relaxation. Section~\ref{sec:algorithm} presents the algorithm, Section~\ref{sec:rounding} analyzes the rounding of a single task, and Section~\ref{sec:closed} computes the work factor $b(a)$ in closed form. Section~\ref{sec:list} studies list scheduling, Section~\ref{sec:general} proves the bound for every $m$, and Section~\ref{sec:opt-a} chooses the parameter $a$. Section~\ref{sec:strict} treats strictly decreasing processing times and concave speedup, Section~\ref{sec:cc13} revisits the strictly decreasing case of~\cite{Chen13}, and Section~\ref{sec:models} checks common speedup models. Section~\ref{sec:limits} discusses the limits of the approach and lower bounds, Section~\ref{sec:numerics} reports numerical checks, and Section~\ref{sec:conclusion} concludes.

\section{Model and Preliminaries}\label{sec:model}

\subsection{The scheduling problem}
We are given tasks $J_1,\dots,J_n$ and a DAG $G=(V,E)$ with $V=\{1,\dots,n\}$. An arc $(i,j)\in E$, also written $i\prec j$, means that $J_j$ may start only after $J_i$ has completed. There are $m$ identical processors. If task $J_j$ is allotted $l\in\{1,\dots,m\}$ processors, its processing time is $p_j(l)>0$ and its \emph{work} is $W_j(l)=l\,p_j(l)$. A schedule assigns to every task an allotment $l_j$ and a start time $S_j$; the task then occupies $l_j$ processors during $[S_j,C_j)$ with $C_j=S_j+p_j(l_j)$, without preemption and without changing its allotment. A schedule is feasible if at every time the running tasks occupy at most $m$ processors in total, and if $S_j\ge C_i$ whenever $i\prec j$. The goal is to minimize the makespan $\Cmax=\max_jC_j$; its optimal value is denoted by $\OPT$.

\subsection{Assumptions}
We use the following conditions, each required for every task $j$.
\begin{itemize}
\item[\Aone] $p_j(1)\ge p_j(2)\ge\dots\ge p_j(m)$: more processors never make a task slower.
\item[\Aonep] $p_j(1)>p_j(2)>\dots>p_j(m)$.
\item[\Aonec] There is an index $l^\circ_j$ such that $p_j(1)>p_j(2)>\dots>p_j(l^\circ_j)=p_j(l^\circ_j+1)=\dots=p_j(m)$: the processing time strictly decreases until it reaches its minimum and stays constant afterwards.
\item[\Atwo] $W_j(1)\le W_j(2)\le\dots\le W_j(m)$: there is no superlinear speedup.
\item[\Athree] The work is a convex function of the processing time; the precise meaning is given in Definition~\ref{def:block}.
\end{itemize}
Clearly \Aonep\ implies \Aonec, which implies \Aone. Conditions \Aone\ and \Atwo\ are the monotonous penalty assumptions~\cite{Blayo99,Lepere02b,Jansen06}, and \Athree\ is the convexity condition of~\cite{Chen13}. Jansen and Zhang~\cite{Jansen11} assume \Aone\ and a concave speedup. Proposition~\ref{prop:concave} shows that a concave speedup is equivalent to \Aonec, \Atwo\ and \Athree\ together; since the LP of~\cite{Jansen11} also needs \Aonep\ (Remark~\ref{rem:jz12}), their setting is \Aonep, \Atwo\ and \Athree.

\subsection{Non-dominated allotments and the envelope work function}
Throughout this subsection we fix one task and drop the index $j$.

\begin{defs}[Non-dominated allotments]\label{def:nondom}
An allotment $l$ is \emph{non-dominated} if $l=1$ or $p(l)<p(l-1)$, and \emph{dominated} otherwise, that is, if $p(l)=p(l-1)$. We list the non-dominated allotments as $1=s_1<s_2<\dots<s_K$.
\end{defs}

\Aone\ gives three immediate consequences.
\begin{itemize}
\item[(a)] If $s_i<l<s_{i+1}$, then $l$ is dominated, so $p(l)=p(l-1)$; by induction, $p(s_i)=p(s_i+1)=\dots=p(s_{i+1}-1)$.
\item[(b)] Since $s_{i+1}$ is non-dominated, $p(s_{i+1})<p(s_{i+1}-1)=p(s_i)$; hence $p(s_1)>p(s_2)>\dots>p(s_K)$.
\item[(c)] All allotments after $s_K$ are dominated, so $p(s_K)=p(s_K+1)=\dots=p(m)$.
\end{itemize}

\begin{fact}\label{fact:dominated}
If $l$ is dominated and $s$ is the largest non-dominated allotment with $s\le l$, then $p(s)=p(l)$ and $W(s)\le W(l)$.
\end{fact}
\begin{proof}
By (a) and (c), $p$ is constant on $\{s,\dots,l\}$, so $p(s)=p(l)$, and $W(s)=s\,p(s)\le l\,p(l)=W(l)$ because $s\le l$.
\end{proof}

\begin{defs}[Blocks, envelope work function, \Athree]\label{def:block}
For $i=1,\dots,K-1$ let $s=s_i$ and $q=s_{i+1}-1$. The set $[s..q]=\{s,\dots,q\}$ is the \emph{$i$-th block}; by (a) and (b), $p(s)=p(s+1)=\dots=p(q)>p(q+1)$. The \emph{envelope work function} $w$ is defined on $[p(m),p(1)]$ as the polygonal line through the points $(p(s_1),W(s_1)),\dots,(p(s_K),W(s_K))$; on the $i$-th segment $[p(q+1),p(s)]$,
\begin{equation}\label{eq:envelope}
w(x)=W(s)+\big(W(q+1)-W(s)\big)\,\frac{p(s)-x}{p(s)-p(q+1)}.
\end{equation}
Let $\sigma_i=\big(W(s_{i+1})-W(s_i)\big)/\big(p(s_i)-p(s_{i+1})\big)$ be the additional work per unit of saved time on the $i$-th segment, which we call the \emph{marginal work} of the segment; the slope of $w$ on that segment is $-\sigma_i$. Condition \Athree\ requires $w$ to be convex. As $x$ increases, the segments are traversed in the order $K-1,K-2,\dots,1$ with slopes $-\sigma_{K-1},\dots,-\sigma_1$, and a polygonal line is convex exactly when its slopes are non-decreasing; hence \Athree\ is equivalent to $\sigma_1\le\sigma_2\le\dots\le\sigma_{K-1}$. If $K=1$, then $p$ is constant and $w$ is defined only at $p(1)$.
\end{defs}

Dominated allotments do not affect the envelope: a dominated point $(p(l),W(l))$ has the same processing time as the non-dominated allotment of its block and at least its work (Fact~\ref{fact:dominated}).

\begin{fact}[Ratio of consecutive processing times]\label{fact:ratio}
Assume \Aone\ and \Atwo. For every block $[s..q]$, the ratio $r=p(q+1)/p(s)$ satisfies $q/(q+1)\le r<1$.
\end{fact}
\begin{proof}
We have $p(q+1)<p(q)=p(s)$, so $r<1$. By \Atwo, $W(q+1)\ge W(q)$, that is, $(q+1)p(q+1)\ge q\,p(q)=q\,p(s)$; dividing by $(q+1)p(s)>0$ gives $r\ge q/(q+1)$.
\end{proof}

\begin{fact}\label{fact:monotone}
Under \Aone\ and \Atwo, the function $w$ is continuous and non-increasing on $[p(m),p(1)]$.
\end{fact}
\begin{proof}
Two consecutive segments both take the value $W(s_i)$ at their common endpoint $p(s_i)$, so $w$ is continuous. By \Atwo, $W(s_{i+1})\ge W(s_i)$, so $\sigma_i\ge0$ and every segment has slope $-\sigma_i\le0$. A continuous polygonal line whose segments are non-increasing is non-increasing.
\end{proof}

\begin{fact}\label{fact:mx}
For every $x\in[p(m),p(1)]$, $w(x)\le m\,x$.
\end{fact}
\begin{proof}
Let $x$ lie on the segment $[p(q+1),p(s)]$ and put $\lambda=(x-p(q+1))/(p(s)-p(q+1))\in[0,1]$. Then $x=\lambda p(s)+(1-\lambda)p(q+1)$ and $(p(s)-x)/(p(s)-p(q+1))=1-\lambda$, so~\eqref{eq:envelope} reads $w(x)=\lambda W(s)+(1-\lambda)W(q+1)$. Since $s,q+1\le m$, we have $W(s)\le m\,p(s)$ and $W(q+1)\le m\,p(q+1)$, hence $w(x)\le m(\lambda p(s)+(1-\lambda)p(q+1))=m\,x$. (If $K=1$, then $x=p(1)$ and $w(x)=W(1)=p(1)\le m\,x$.)
\end{proof}

\begin{fact}[Scaling]\label{fact:scaling}
Multiplying all processing times of a task by a constant $t>0$ multiplies $W$ and $w$ by $t$, leaves every comparison of the form ``$p(s)\le a\,x$'' unchanged when $x$ is scaled as well, and leaves all ratios of works unchanged. Hence, when we study a single block $[s..q]$, we may assume $p(s)=1$. Then $p(q+1)=r$, $W(s)=s$, $W(q+1)=(q+1)r$, and on $[r,1]$
\begin{equation}\label{eq:normalized}
w(x)=s+\big((q+1)r-s\big)\frac{1-x}{1-r}.
\end{equation}
\end{fact}
\begin{proof}
The first statements follow from $W(l)=l\,p(l)$ and~\eqref{eq:envelope}; \eqref{eq:normalized} is~\eqref{eq:envelope} with $p(s)=1$ and $p(q+1)=r$.
\end{proof}

\subsection{The linear programming relaxation}
For every task $j$, let $1=s_{j,1}<\dots<s_{j,K_j}$ be its non-dominated allotments and $\sigma_{j,1},\dots,\sigma_{j,K_j-1}$ its marginal works (Definition~\ref{def:block}); the slope of the envelope $w_j$ on its $i$-th segment is $-\sigma_{j,i}$. The variables are a fractional processing time $x_j$, a work $w_j$ and a completion time $C_j$ for every task, together with $L$ and $C$:
\begin{align}
\min\ & C \notag\\
\text{s.t.}\ & C_i+x_j\le C_j && \forall (i,j)\in E, \label{lp:prec}\\
& x_j\le C_j\le L && \forall j, \label{lp:start}\\
& L\le C,\qquad \frac1m\sum_{j=1}^n w_j\le C, && \label{lp:obj}\\
& p_j(m)\le x_j\le p_j(1) && \forall j, \label{lp:range}\\
& w_j\ge W_j(1),\quad w_j\ge W_j(s_{j,i})+\sigma_{j,i}\big(p_j(s_{j,i})-x_j\big) && \forall j,\ i=1,\dots,K_j-1. \label{lp:work}
\end{align}
We call this LP $(\mathrm{P})$. It has $3n+2$ variables and $O(nm+|E|)$ constraints, so it can be solved in polynomial time. We denote an optimal solution by $x_j^*,w_j^*,C_j^*,L^*$ and its value by $C^*$.

\begin{lem}\label{lem:epigraph}
Assume \Aone, \Atwo\ and \Athree, and fix a task (dropping $j$). For $x\in[p(m),p(1)]$ let $\ell_i(x)=W(s_i)+\sigma_i(p(s_i)-x)$. Then $\max\{W(1),\ell_1(x),\dots,\ell_{K-1}(x)\}=w(x)$. Consequently, constraint~\eqref{lp:work} is equivalent to $w_j\ge w_j(x_j)$.
\end{lem}
\begin{proof}
(a) If $x$ lies on the $k$-th segment, then $\ell_k(x)=w(x)$: both $\ell_k$ and $w$ restricted to that segment are affine, both take the value $W(s_k)$ at $p(s_k)$, and both have slope $-\sigma_k$.

(b) For every $i$ and $x$, $\ell_i(x)\le w(x)$. If $x$ lies on the $i$-th segment this is (a). Suppose $x<p(s_{i+1})$, that is, $x$ lies to the left of the $i$-th segment. At $p(s_{i+1})$ both functions take the value $W(s_{i+1})$, because $\ell_i(p(s_{i+1}))=W(s_i)+\sigma_i(p(s_i)-p(s_{i+1}))=W(s_{i+1})$. Moving left from $p(s_{i+1})$ to $x$, the function $w$ traverses the segments $i+1,i+2,\dots$, whose marginal works satisfy $\sigma_k\ge\sigma_i$ by \Athree; hence $w$ increases at rate at least $\sigma_i$, while $\ell_i$ increases at rate exactly $\sigma_i$, and $w(x)\ge\ell_i(x)$. Suppose $x>p(s_i)$. At $p(s_i)$ both functions equal $W(s_i)$. Moving right, $w$ traverses segments with marginal works $\sigma_k\le\sigma_i$ and decreases at rate at most $\sigma_i$, while $\ell_i$ decreases at rate exactly $\sigma_i$; again $w(x)\ge\ell_i(x)$.

(c) $W(1)=w(p(1))\le w(x)$ by Fact~\ref{fact:monotone}.

By (b) and (c) the maximum is at most $w(x)$, and by (a) it is at least $w(x)$. If $K=1$, then $x=p(1)$ and the maximum is $W(1)=w(x)$.
\end{proof}

\begin{lem}[Relaxation]\label{lem:relax}
Under \Aone, \Atwo\ and \Athree, $C^*\le\OPT$.
\end{lem}
\begin{proof}
Take an optimal schedule with allotments $l_j^\circ$, start times $S_j^\circ$ and completion times $C_j^\circ$. Set $x_j=p_j(l_j^\circ)$, $w_j=W_j(l_j^\circ)$, $C_j=C_j^\circ$ and $L=C=\OPT$. We check the constraints.
Constraint~\eqref{lp:range} holds by \Aone.
For~\eqref{lp:work} it suffices, by Lemma~\ref{lem:epigraph}, to show $w_j\ge w_j(x_j)$. If $l_j^\circ$ is non-dominated, then $w_j(x_j)=W_j(l_j^\circ)=w_j$. Otherwise let $s$ be the non-dominated allotment of Fact~\ref{fact:dominated}; then $p_j(s)=x_j$ and $w_j(x_j)=W_j(s)\le W_j(l_j^\circ)=w_j$.
For~\eqref{lp:prec}, $J_j$ starts after $J_i$ completes, so $C_j^\circ=S_j^\circ+x_j\ge C_i^\circ+x_j$.
For~\eqref{lp:start}, $C_j^\circ=S_j^\circ+x_j\ge x_j$ and $C_j^\circ\le\OPT=L$.
For~\eqref{lp:obj}, $L=C$, and $\sum_jw_j=\sum_jl_j^\circ p_j(l_j^\circ)$ is the integral over $[0,\OPT]$ of the number of busy processors, which is at most $m$; hence $\sum_jw_j\le m\,\OPT=m\,C$.
The solution is feasible with value $\OPT$, so $C^*\le\OPT$.
\end{proof}

\begin{lem}[Paths and work]\label{lem:paths}
For a task $J$ write $x^*(J)$ and $C^*(J)$ for its values in the optimal LP solution. If $P_1\prec P_2\prec\dots\prec P_k$ is a path of $G$, then $\sum_{i=1}^kx^*(P_i)\le C^*$. Moreover, $\sum_jw_j^*\le m\,C^*$, and under \Athree\ $w_j^*\ge w_j(x_j^*)$ for every $j$.
\end{lem}
\begin{proof}
By~\eqref{lp:start}, $C^*(P_1)\ge x^*(P_1)$, and by~\eqref{lp:prec}, $C^*(P_i)\ge C^*(P_{i-1})+x^*(P_i)$ for $i=2,\dots,k$. Adding these $k$ inequalities, the intermediate terms cancel and $C^*(P_k)\ge\sum_ix^*(P_i)$. By~\eqref{lp:start} and~\eqref{lp:obj}, $C^*\ge L^*\ge C^*(P_k)$. The second statement is~\eqref{lp:obj}, and the third follows from~\eqref{lp:work} and Lemma~\ref{lem:epigraph}.
\end{proof}

\begin{rem}[Relation to earlier LPs]\label{rem:lp-relation}
The LP of~\cite{Chen13} describes the work of a task through virtual tasks, one for each segment of the envelope, following~\cite{Jansen06}. After the adjustment of~\cite[Lemma~3.1]{Chen13}, which does not increase the work, the virtual tasks of $J_j$ contribute exactly $w_j(x_j)-W_j(1)$, so the total LP work equals $\sum_jw_j(x_j)$. Under \Athree, $(\mathrm{P})$ is therefore the same relaxation written in epigraph form, with one difference: we include the constraint $x_j\le C_j$ in~\eqref{lp:start}. As printed, the LP of~\cite{Chen13} bounds a completion time from below only by $0$ and by the completion times of predecessors, so for a task without predecessors it does not force $C_j\ge x_j$; the bound $\sum_ix^*(P_i)\le C^*$ of Lemma~\ref{lem:paths} needs this constraint, and Lemma~\ref{lem:relax} shows that it keeps the LP a relaxation. The LPs~(7) and~(9) of~\cite{Jansen11} contain the same constraints $0\le C_j\le L$ and also omit $x_j\le C_j$. Without this constraint an LP of this form still bounds $\OPT$ from below, but it no longer bounds the length of a path: for a single task with $p(1)=\dots=p(m)$ its value is $p(1)/m$, while $x_1^*=\OPT=p(1)$, so Lemma~\ref{lem:paths} fails and the integrality gap is at least $m$.
\end{rem}

\begin{rem}[Without convexity]\label{rem:hull}
Without \Athree, Lemma~\ref{lem:epigraph}(b) fails, and~\eqref{lp:work} must be replaced by the lower convex hull of the points $\{(p(l),W(l))\}$ to keep $(\mathrm{P})$ a relaxation. Section~\ref{sec:limits} shows that this loses the advantage of speedup-aware rounding.
\end{rem}

\section{The Algorithm}\label{sec:algorithm}
Algorithm $\SAR(a,\mu)$ has two parameters: the rounding parameter $a>1$ and the cap $\mu$, an integer with $1\le\mu\le(m+1)/2$. The main theorems use $\mu=\lceil\beta(a)m\rceil$ with $\beta(a)=2/(a+2+\sqrt{a^2+4})$. Only the rounding step differs from the two-phase algorithms of~\cite{Jansen06,Chen13}.

\begin{algorithm}[h]
\caption{$\SAR(a,\mu)$}\label{alg:sar}
\begin{algorithmic}[1]
\REQUIRE tasks with processing times $p_j(\cdot)$, DAG $G$, $m$, parameters $a>1$ and $\mu$
\ENSURE a feasible schedule
\STATE Solve $(\mathrm{P})$ and let $x_j^*$ be the optimal fractional processing times.
\FOR{every task $j$}
  \IF{$x_j^*=p_j(s)$ for a non-dominated allotment $s$}
    \STATE $l_j'\leftarrow s$
  \ELSE
    \STATE let $[s..q]$ be the block with $p_j(q+1)<x_j^*<p_j(s)$
    \IF{$p_j(s)\le a\,x_j^*$}
      \STATE $l_j'\leftarrow s$ \COMMENT{round up: fewer processors, longer time}
    \ELSE
      \STATE $l_j'\leftarrow q+1$ \COMMENT{round down: more processors, shorter time}
    \ENDIF
  \ENDIF
  \STATE $l_j\leftarrow\min\{l_j',\mu\}$
\ENDFOR
\STATE Run \LIST\ with the allotments $l_j$.
\end{algorithmic}
\end{algorithm}

A task with $l_j'>\mu$ is called \emph{capped}. In floating-point arithmetic the test $x_j^*=p_j(s)$ in line~3 is made with a small tolerance; the ancillary implementation uses $10^{-12}$. \LIST\ is the following event-driven list scheduling procedure. Fix any priority order of the tasks. The \emph{decision times} are $0$ and the completion times of the tasks. At each decision time, \LIST\ scans, in priority order, the tasks that are ready (all predecessors have completed) and have not started yet, and starts each scanned task for which at least $l_j$ processors are idle at that moment; then it advances to the next decision time. Section~\ref{sec:list} also covers the variant used in~\cite{Jansen06,Chen13}, which repeatedly places the ready task with the earliest feasible start time.

\begin{lem}[Well-definedness]\label{lem:welldef}
For every task, the rounding step falls into exactly one case. \LIST\ terminates and returns a feasible schedule.
\end{lem}
\begin{proof}
By~\eqref{lp:range}, $x_j^*\in[p_j(m),p_j(1)]=[p_j(s_K),p_j(s_1)]$. The points $p_j(s_K)<\dots<p_j(s_1)$ split this interval into $K$ points and $K-1$ open intervals, so $x_j^*$ either equals some $p_j(s_i)$ or lies in exactly one open interval $(p_j(s_{i+1}),p_j(s_i))$, which corresponds to the block $[s_i..s_{i+1}-1]$.
\LIST\ starts a task only if enough processors are idle and only if all its predecessors have completed, so the schedule is feasible. Suppose that after a scan no task is running but some task is unscheduled. An unscheduled task without unscheduled predecessors exists because $G$ is acyclic; all its predecessors have started and, since nothing is running, completed. The task is ready, and $m\ge\mu\ge l_j$ processors are idle, so the scan would have started it, a contradiction. Hence after each scan either all tasks are scheduled or some task is running and a next decision time exists. Every task starts once, so there are at most $n+1$ decision times.
\end{proof}

\begin{rem}[Large $a$]\label{rem:large-a}
If $a\ge2$, every task is rounded up. Indeed, for every block and every $x^*\in(p(q+1),p(s))$, Fact~\ref{fact:ratio} gives $p(s)=p(q+1)/r\le\frac{q+1}{q}p(q+1)\le2p(q+1)<2x^*\le a\,x^*$. In this case $\SAR$ coincides with the rounding of~\cite{Chen13} for $\rho=0$, except when $x^*=p(s_i)$ for a non-dominated allotment $s_i$ with $i\ge2$: then $\SAR$ keeps $s_i$, whereas the rule of~\cite{Chen13} as printed, with its strict inequality and $\rho=0$, moves to $s_{i-1}$. This does not affect any bound, because $p(s_{i-1})\le2p(s_i)$ by Fact~\ref{fact:ratio} and $W(s_{i-1})\le W(s_i)=w(x^*)$ by \Atwo.
\end{rem}

\begin{rem}[Threshold rounding]\label{rem:threshold}
The rules of~\cite{Jansen06,Chen13} use the relative position $\lambda=(x^*-p(q+1))/(p(s)-p(q+1))$ of $x^*$ in its block and round up if and only if $\lambda\ge\rho$. When a task is rounded down, the work can grow by the factor $1/(1-\rho)$, because $w(x^*)=\lambda W(s)+(1-\lambda)W(q+1)\ge(1-\lambda)W(q+1)$ (see the proof of Fact~\ref{fact:mx}), and this factor is approached when $W(s)$ is much smaller than $W(q+1)$, for instance for a block $[1..q]$ with large $q$. When it is rounded up, $x^*\ge\rho\,p(s)+(1-\rho)p(q+1)$, and the time factor depends on $\rho$ and on the ratio $r=p(q+1)/p(s)$. Speedup-aware rounding compares $p(s)$ with $a\,x^*$ directly: the time factor $a$ holds by construction, and Fact~\ref{fact:ratio} keeps the work factor small, as shown in Sections~\ref{sec:rounding} and~\ref{sec:closed}. Equivalently, with $p(s)=1$ and $r=p(q+1)$ we have $x^*=r+\lambda(1-r)$, so speedup-aware rounding rounds up exactly when $\lambda\ge\rho(r):=\max\{0,(\alpha-r)/(1-r)\}$, where $\alpha=1/a$. It is thus a threshold rule whose threshold depends on the ratio $r$ of the block: a block with a large gain (small $r$) gets a high threshold, and a block with $r\ge\alpha$ is always rounded up.
\end{rem}

\textbf{Running time.} The LP is solved in polynomial time. Computing the blocks, the rounding and the caps takes $O(nm)$ time. \LIST\ has at most $n+1$ decision times; each scan takes $O(n)$ time, and maintaining the numbers of uncompleted predecessors takes $O(|E|)$ time in total, so \LIST\ runs in $O(n^2+|E|)$ time.

\section{Rounding a Single Task}\label{sec:rounding}
We show that every task, after rounding, satisfies three properties: its processing time is at most $a\,x^*$, its work is at most $b(a)\,w(x^*)$, and its work is at most $m\,x^*$. This section defines $b(a)$ and proves these properties; Section~\ref{sec:closed} computes $b(a)$.

\begin{defs}\label{def:F}
Let $\alpha=1/a<1$. For integers $1\le s\le q$ and a real number $r$ put
\[
N(r)=(\alpha-r)\big((q+1)r-s\big),\qquad D(r)=s\alpha+\big(q+1-s-(q+1)\alpha\big)r,\qquad F(q,s;r)=\frac{N(r)}{D(r)}.
\]
The triple $(q,s,r)$ is \emph{feasible} if $q/(q+1)\le r<\alpha$. Let $b(a)=1+\sup F(q,s;r)$, where the supremum is over all feasible triples, and $b(a)=1$ if no triple is feasible.
\end{defs}

\textbf{Origin and meaning of $F$.} Rounding up never increases the work: if $s$ is chosen, then $W(s)=w(p(s))\le w(x^*)$ by Fact~\ref{fact:monotone}. The quantities of Definition~\ref{def:F} describe what rounding down can cost. Consider a block $[s..q]$, scaled by Fact~\ref{fact:scaling} so that $p(s)=1$, and let $r=p(q+1)$, so that $W(s)=s$ and $W(q+1)=(q+1)r$. By~\eqref{eq:normalized}, the envelope on $[r,1]$ is the segment from $(r,W(q+1))$ to $(1,W(s))$:
\[
w(x)=W(q+1)-\frac{x-r}{1-r}\big(W(q+1)-W(s)\big),\qquad r\le x\le1.
\]
For $x^*\in(r,1)$, speedup-aware rounding keeps the slower allotment $s$ exactly when $1\le a\,x^*$, that is, when $x^*\ge\alpha$; otherwise it rounds down to $q+1$ and replaces the envelope work $w(x^*)$ by $W(q+1)$. Rounding down therefore happens exactly for $x^*\in(r,\alpha)$. This interval is non-empty only if $r<\alpha$, the upper bound in the definition of a feasible triple; if $r\ge\alpha$, the task is always rounded up, at no extra work. The lower bound $r\ge q/(q+1)$ is Fact~\ref{fact:ratio}. Since $w$ is non-increasing, the ratio $W(q+1)/w(x^*)$ does not decrease as $x^*$ grows, so the extra work is largest when $x^*$ approaches the threshold $\alpha$ from below. At the threshold, the formula above and the identity $(1-r)\,w(\alpha)=D(r)$ (Step~3 of the proof of Lemma~\ref{lem:rounding}) give
\[
W(q+1)-w(\alpha)=\frac{(\alpha-r)\big((q+1)r-s\big)}{1-r}=\frac{N(r)}{1-r},\qquad w(\alpha)=\frac{D(r)}{1-r},
\]
and hence
\[
F(q,s;r)=\frac{N(r)}{D(r)}=\frac{W(q+1)-w(\alpha)}{w(\alpha)}.
\]
Thus $F(q,s;r)$ is the relative extra work caused by rounding down a task whose block has the parameters $q$, $s$ and $r$ and whose fractional processing time lies just below the threshold, and $1+F(q,s;r)=W(q+1)/w(\alpha)$ is the corresponding work factor (Remark~\ref{rem:worst}). The numerator $N(r)$ is the product of two quantities: the distance $\alpha-r$ between the threshold and the processing time of the faster allotment, and the work difference $W(q+1)-W(s)=(q+1)r-s$ across the block. The denominator $D(r)$ is the envelope work at the threshold multiplied by the length $1-r$ of the segment. As $r\to\alpha$, the interval $(r,\alpha)$ shrinks and $F\to0$. A larger $s$ means more work at the slow end of the block and hence a smaller relative increase (Lemma~\ref{lem:F-s}). A larger $q$ makes the work difference $(q+1)r-s\ge q-s$ larger, but it also forces $r\ge q/(q+1)$ closer to $1$, so that only blocks with $q<1/(a-1)$ can be rounded down at all (Lemma~\ref{lem:small-q}). The value $b(a)=1+\sup F$ is therefore the supremum of the work factors that rounding down can cause, over all blocks; Lemma~\ref{lem:rounding} shows that it bounds the work factor of every task, and Remark~\ref{rem:worst} that it is attained in the limit.

\begin{lem}\label{lem:F-sign}
For every feasible triple, $D(r)>0$ and $N(r)\ge0$. Hence $F(q,s;r)\ge0$ and $b(a)\ge1$.
\end{lem}
\begin{proof}
Expanding, $D(r)=s\alpha-sr+(q+1)r-(q+1)\alpha r=s(\alpha-r)+(q+1)(1-\alpha)r$. Since $r<\alpha$, the first term is positive; since $\alpha<1$ and $r>0$, the second term is positive. In $N(r)$, the first factor $\alpha-r$ is positive and the second factor satisfies $(q+1)r-s\ge q-s\ge0$ because $r\ge q/(q+1)$ and $s\le q$.
\end{proof}

\begin{lem}[Rounding lemma]\label{lem:rounding}
Assume \Aone, \Atwo\ and \Athree. For every task, the allotment $l'$ chosen by the rounding step satisfies
\begin{itemize}
\item[(i)] $p(l')\le a\,x^*$;
\item[(ii)] $W(l')\le b(a)\,w(x^*)$;
\item[(iii)] $W(l')\le m\,x^*$.
\end{itemize}
\end{lem}
\begin{proof}
We follow the three cases of the rounding step.

\emph{Case 0: $x^*=p(s_i)$ and $l'=s_i$.} (i) $p(l')=x^*<a\,x^*$. (ii) $W(l')=w(x^*)\le b(a)\,w(x^*)$ because $b(a)\ge1$. (iii) $W(l')=w(x^*)\le m\,x^*$ by Fact~\ref{fact:mx}.

\emph{Case 1: $p(q+1)<x^*<p(s)$, $p(s)\le a\,x^*$ and $l'=s$.} (i) is the rounding condition. (ii) Since $x^*<p(s)$, Fact~\ref{fact:monotone} gives $W(s)=w(p(s))\le w(x^*)\le b(a)\,w(x^*)$. (iii) $W(s)\le w(x^*)\le m\,x^*$ by Fact~\ref{fact:mx}.

\emph{Case 2: $p(q+1)<x^*<p(s)$, $p(s)>a\,x^*$ and $l'=q+1$.} (i) $p(q+1)<x^*<a\,x^*$. (iii) $W(q+1)=(q+1)p(q+1)\le m\,p(q+1)<m\,x^*$ because $q+1\le m$. For (ii) we proceed in four steps.
\begin{enumerate}
\item By Fact~\ref{fact:scaling} we may assume $p(s)=1$ and $p(q+1)=r$. The rounding condition $1>a\,x^*$ means $x^*<\alpha$, and $x^*>r$ gives $r<\alpha$. Together with Fact~\ref{fact:ratio}, $(q,s,r)$ is feasible.
\item Since $r<\alpha<1$, the point $\alpha$ lies in $[r,1]$. As $x^*<\alpha$, Fact~\ref{fact:monotone} gives $w(x^*)\ge w(\alpha)$, and $w(\alpha)>0$ as a convex combination of two positive works. Hence $W(q+1)/w(x^*)\le(q+1)r/w(\alpha)$.
\item By~\eqref{eq:normalized},
$(1-r)\,w(\alpha)=s(1-r)+\big((q+1)r-s\big)(1-\alpha)=s\alpha+\big(q+1-s-(q+1)\alpha\big)r=D(r)$.
\item Further,
$(q+1)r(1-r)-D(r)=(q+1)r-(q+1)r^2-s\alpha-(q+1)r+sr+(q+1)\alpha r=-(q+1)r(r-\alpha)+s(r-\alpha)=N(r)$.
\end{enumerate}
Therefore $(q+1)r/w(\alpha)=(q+1)r(1-r)/D(r)=(D(r)+N(r))/D(r)=1+F(q,s;r)\le b(a)$, that is, $W(q+1)\le b(a)\,w(x^*)$.
\end{proof}

\begin{rem}[The worst case]\label{rem:worst}
In Case~2, $w$ is continuous on $[r,1]$, so $W(q+1)/w(x^*)\to(q+1)r/w(\alpha)=1+F(q,s;r)$ as $x^*\to\alpha^-$. Thus $b(a)$ is exactly the worst-case work factor of speedup-aware rounding; Example~\ref{ex:tight} gives instances that attain it.
\end{rem}

\begin{coro}[Capped tasks]\label{coro:capped}
Assume \Aone, \Atwo\ and \Athree. If $l'>\mu$, then the capped allotment $l=\mu$ satisfies $p(\mu)\le(m/\mu)\,x^*$ and $W(\mu)\le b(a)\,w(x^*)$.
\end{coro}
\begin{proof}
Since $\mu<l'$, \Atwo\ gives $W(\mu)\le W(l')$. Hence $p(\mu)=W(\mu)/\mu\le W(l')/\mu\le m\,x^*/\mu$ by Lemma~\ref{lem:rounding}(iii), and $W(\mu)\le W(l')\le b(a)\,w(x^*)$ by Lemma~\ref{lem:rounding}(ii).
\end{proof}

\section{The Work Factor in Closed Form}\label{sec:closed}
The definition of $b(a)$ takes a supremum over infinitely many pairs $(q,s)$. We first show that only blocks with $q<1/(a-1)$ can be rounded down, that $s=1$ is the worst choice for fixed $q$, and that among blocks with $s=q$ the block $(1,1)$ is the worst. Then we evaluate the two remaining candidates. Throughout, $\alpha=1/a$ and $u=\sqrt{a-1}$, so that $a=1+u^2$.

\begin{lem}[Only small blocks are rounded down]\label{lem:small-q}
A feasible $r$ exists for $q$ if and only if $q<1/(a-1)$. Consequently $b(a)=1$ for $a\ge2$; for $3/2\le a<2$ only $q=1$ is feasible; and for $4/3\le a<3/2$ only $q\in\{1,2\}$ is feasible. Moreover, $b(a)<\infty$ for every $a>1$.
\end{lem}
\begin{proof}
The interval $[q/(q+1),\alpha)$ is non-empty if and only if $q/(q+1)<1/a$, that is, $aq<q+1$, or $q(a-1)<1$. For $a\ge2$ we have $1/(a-1)\le1$, and no positive integer is feasible. For $3/2\le a<2$, $1<1/(a-1)\le2$, so only $q=1$; for $4/3\le a<3/2$, $2<1/(a-1)\le3$, so only $q\in\{1,2\}$. For every $a>1$ there are finitely many feasible pairs $(q,s)$, and for each of them $D>0$ on the compact interval $[q/(q+1),\alpha]$ (the proof of Lemma~\ref{lem:F-sign} works also for $r=\alpha$, where $D(\alpha)=(q+1)(1-\alpha)\alpha>0$). Hence $F(q,s;\cdot)$ is continuous and bounded there, and $b(a)<\infty$.
\end{proof}

\begin{lem}[$F$ decreases in $s$]\label{lem:F-s}
If $(q,s,r)$ and $(q,s',r)$ are feasible and $s<s'$, then $F(q,s';r)\le F(q,s;r)$. In particular $F(q,s;r)\le F(q,1;r)$.
\end{lem}
\begin{proof}
Feasibility depends only on $q$ and $r$. By the proof of Lemma~\ref{lem:F-sign}, $F=(\alpha-r)\big((q+1)r-s\big)/\big(s(\alpha-r)+(q+1)(1-\alpha)r\big)$. When $s$ increases, the numerator does not increase and stays non-negative, because $\alpha-r>0$ is fixed and $(q+1)r-s$ decreases but stays $\ge0$ (as $(q+1)r\ge q\ge s'$); the denominator increases and stays positive, because $s(\alpha-r)$ increases. Hence the ratio does not increase.
\end{proof}

\begin{lem}[Critical points]\label{lem:critical}
Let $\kappa=q+1-s-(q+1)\alpha$, so that $D(r)=s\alpha+\kappa r$. Then $\frac{d}{dr}F(q,s;r)$ has the same sign as $-(\kappa r^2+2s\alpha r-s\alpha)$.
\end{lem}
\begin{proof}
Expanding, $N(r)=-(q+1)r^2+\big((q+1)\alpha+s\big)r-s\alpha$, so $N'(r)=-2(q+1)r+(q+1)\alpha+s$ and $D'(r)=\kappa$. Term by term,
\begin{align*}
N'D&=-2(q+1)s\alpha\,r-2(q+1)\kappa\,r^2+\big((q+1)\alpha+s\big)s\alpha+\big((q+1)\alpha+s\big)\kappa\,r,\\
ND'&=-(q+1)\kappa\,r^2+\big((q+1)\alpha+s\big)\kappa\,r-s\alpha\kappa.
\end{align*}
In the difference $N'D-ND'$, the coefficient of $r^2$ is $-(q+1)\kappa$, the coefficient of $r$ is $-2(q+1)s\alpha$ (the terms with $\kappa$ cancel), and the constant term is $s\alpha\big((q+1)\alpha+s+\kappa\big)=s\alpha(q+1)$ because $(q+1)\alpha+s+\kappa=q+1$. Hence $N'D-ND'=-(q+1)(\kappa r^2+2s\alpha r-s\alpha)$. Since $F'=(N'D-ND')/D^2$ with $D^2>0$ and $q+1>0$, the claim follows.
\end{proof}

\begin{lem}[Substitution for $s=q$]\label{lem:subst}
Let $s=q$, $c_q=(q+1)\alpha-q$ and $\delta=(q+1)r-q$, the additional work caused by rounding down. The map $r\mapsto\delta$ is a bijection from $[q/(q+1),\alpha)$ onto $[0,c_q)$, and
\[
F(q,q;r)=G(q;\delta):=\frac{\delta(c_q-\delta)}{q-(q-1+c_q)\,\delta}.
\]
\end{lem}
\begin{proof}
The map is affine and strictly increasing, with $\delta=0$ at $r=q/(q+1)$ and $\delta=c_q$ at $r=\alpha$. Substituting $r=(q+\delta)/(q+1)$: $\alpha-r=(c_q-\delta)/(q+1)$ and $(q+1)r-q=\delta$, so $N=\delta(c_q-\delta)/(q+1)$. Further, $(q+1)(1-\alpha)=1-c_q$, so $(q+1)(1-\alpha)r=(1-c_q)(q+\delta)/(q+1)$. By the proof of Lemma~\ref{lem:F-sign}, $D=q(\alpha-r)+(q+1)(1-\alpha)r$, hence
\[
(q+1)D=q(c_q-\delta)+(1-c_q)(q+\delta)=q-(q-1+c_q)\delta.
\]
Dividing $N$ by $D$ gives the formula.
\end{proof}

\begin{lem}[Among blocks with $s=q$, the block $(1,1)$ is the worst]\label{lem:singleton}
For every $q\ge1$ and $\delta\in[0,c_q)$, $G(q;\delta)\le G(1;\delta)$. Consequently $\sup_rF(q,q;r)\le\sup_rF(1,1;r)$.
\end{lem}
\begin{proof}
First, $c_1-c_q=(2\alpha-1)-\big((q+1)\alpha-q\big)=(q-1)(1-\alpha)\ge0$, so $[0,c_q)\subseteq[0,c_1)$ and $G(1;\delta)$ is defined.
Numerators: $0\le\delta(c_q-\delta)\le\delta(c_1-\delta)$.
Denominators: let $D_q=q-(q-1+c_q)\delta$ and $D_1=1-c_1\delta$. Since $q-1+c_q-c_1=(q-1)-(q-1)(1-\alpha)=(q-1)\alpha$, we get $D_q-D_1=(q-1)-(q-1)\alpha\delta=(q-1)(1-\alpha\delta)\ge0$ because $\alpha<1$ and $\delta<1$. Moreover $\delta<c_1<1$ gives $D_1>1-c_1^2>0$. Hence $D_q\ge D_1>0$.
A smaller non-negative numerator and a larger positive denominator give $G(q;\delta)\le G(1;\delta)$. The second statement follows from Lemma~\ref{lem:subst}, since $\sup_rF(q,q;r)$ is the supremum of $G(q;\cdot)$ over $[0,c_q)$, which is at most the supremum of $G(1;\cdot)$ over $[0,c_1)$.
\end{proof}

\begin{prop}[The block $(1,1)$]\label{prop:11}
For $1<a<2$, $\sup_rF(1,1;r)=\big((1-u)/(1+u)\big)^2$, attained at $r=1/(1+u)$.
\end{prop}
\begin{proof}
For $q=s=1$, Lemma~\ref{lem:subst} gives $F(1,1;r)=G(1;\delta)=\delta(c-\delta)/(1-c\delta)$ with $\delta=2r-1\in[0,c)$ and $c=c_1=2\alpha-1$. Since $a=1+u^2$, $c=2/(1+u^2)-1=(1-u^2)/(1+u^2)\in(0,1)$.
The derivative is $G'(\delta)=\big[(c-2\delta)(1-c\delta)+c\delta(c-\delta)\big]/(1-c\delta)^2$, and the numerator expands to $c-c^2\delta-2\delta+2c\delta^2+c^2\delta-c\delta^2=c-2\delta+c\delta^2$. The equation $c\delta^2-2\delta+c=0$ has the roots $(1\pm\sqrt{1-c^2})/c$. The larger root exceeds $1/c>1>c$ and lies outside the interval; the smaller root $\delta_0=(1-\sqrt{1-c^2})/c=c/(1+\sqrt{1-c^2})$ lies in $(0,c)$. Since $G$ is continuous on $[0,c]$, vanishes at both endpoints and is positive inside, its maximum is attained at the unique critical point $\delta_0$.
Now $(1+u^2)^2-(1-u^2)^2=4u^2$, so $\sqrt{1-c^2}=2u/(1+u^2)$, $1+\sqrt{1-c^2}=(1+u)^2/(1+u^2)$ and $\delta_0=(1-u^2)/(1+u)^2=(1-u)/(1+u)$. From $c\delta_0^2-2\delta_0+c=0$ we get $c=2\delta_0/(1+\delta_0^2)$, hence $c-\delta_0=\delta_0(1-\delta_0^2)/(1+\delta_0^2)$ and $1-c\delta_0=(1-\delta_0^2)/(1+\delta_0^2)$, and $G(1;\delta_0)=\delta_0^2$. The corresponding ratio is $r=(1+\delta_0)/2=1/(1+u)$, which lies in $[1/2,\alpha)$ because $u<1$.
\end{proof}

\begin{prop}[The block $(2,1)$]\label{prop:21}
\textup{(a)} For $1<a<3/2$, the ratio $r=2/3$ is feasible for $(q,s)=(2,1)$ and $F(2,1;2/3)=(3-2a)/(4a-3)$; hence $b(a)\ge2a/(4a-3)$.
\textup{(b)} For $9/8<a<3/2$, $F(2,1;\cdot)$ is strictly decreasing on $[2/3,\alpha)$, so $\sup_rF(2,1;r)=(3-2a)/(4a-3)$.
\end{prop}
\begin{proof}
(a) Here $q/(q+1)=2/3$, and $a<3/2$ is equivalent to $\alpha>2/3$, so $r=2/3$ is feasible. From the definitions, $N(2/3)=(\alpha-2/3)(3\cdot\frac23-1)=\alpha-2/3$, $\kappa=3-1-3\alpha=2-3\alpha$ and $D(2/3)=\alpha+(2-3\alpha)\frac23=\frac43-\alpha$. Hence $F=(3\alpha-2)/(4-3\alpha)$, which equals $(3-2a)/(4a-3)$ after multiplying numerator and denominator by $a$. Finally $b(a)\ge1+(3-2a)/(4a-3)=2a/(4a-3)$.
(b) By Lemma~\ref{lem:critical} with $s=1$, $F'$ has the sign of $-\psi(r)$ with $\psi(r)=\kappa r^2+2\alpha r-\alpha$. Now $\alpha\in(2/3,8/9)$ and $\kappa=2-3\alpha<0$, so $\psi$ is a concave quadratic and its minimum over $[2/3,\alpha]$ is attained at an endpoint. We have $\psi(2/3)=\frac49(2-3\alpha)+\frac43\alpha-\alpha=\frac89-\alpha>0$ because $\alpha<8/9$, and $\psi(\alpha)=(2-3\alpha)\alpha^2+2\alpha^2-\alpha=-\alpha(3\alpha^2-4\alpha+1)=\alpha(1-\alpha)(3\alpha-1)>0$. Hence $\psi>0$ and $F'<0$.
\end{proof}

\begin{lem}[Crossing identity]\label{lem:crossing}
For $1<a<3/2$,
\[
\Big(\frac{1-u}{1+u}\Big)^2-\frac{3-2a}{4a-3}=\frac{2u(3u-2)(u^2+1)}{(1+u)^2(1+4u^2)}.
\]
Hence the left-hand side is non-negative for $a\ge13/9$ and negative for $a<13/9$.
\end{lem}
\begin{proof}
With $a=1+u^2$ we have $3-2a=1-2u^2$ and $4a-3=1+4u^2$. Over the common denominator $(1+u)^2(1+4u^2)$, the numerator is $(1-u)^2(1+4u^2)-(1-2u^2)(1+u)^2$. Here
$(1-u)^2(1+4u^2)=1-2u+5u^2-8u^3+4u^4$ and $(1-2u^2)(1+u)^2=1+2u-u^2-4u^3-2u^4$, whose difference is $-4u+6u^2-4u^3+6u^4=2u(3u^3-2u^2+3u-2)=2u(3u-2)(u^2+1)$. The denominator is positive, $u>0$ and $u^2+1>0$, so the sign is that of $3u-2$; and $u\ge2/3$ if and only if $a\ge13/9$.
\end{proof}

\begin{thm}[Closed form]\label{thm:closed}
For $13/9\le a<2$,
\[
b(a)=1+g(a),\qquad g(a)=\Big(\frac{1-\sqrt{a-1}}{1+\sqrt{a-1}}\Big)^2.
\]
In particular $b(13/9)=26/25$ and $b(3/2)=18-12\sqrt2$.
\end{thm}
\begin{proof}
By Lemma~\ref{lem:small-q}, only blocks with $q\le2$ are feasible (and only $q=1$ when $a\ge3/2$), so the candidates are $(q,s)\in\{(1,1),(2,1),(2,2)\}$. By Lemma~\ref{lem:F-s}, $F(2,2;r)\le F(2,1;r)$, so it suffices to compare $(1,1)$ with $(2,1)$. For $3/2\le a<2$ only $(1,1)$ remains, and Proposition~\ref{prop:11} gives the claim. For $13/9\le a<3/2$, Proposition~\ref{prop:21}(b) gives $\sup_rF(2,1;r)=(3-2a)/(4a-3)$, which is at most $\sup_rF(1,1;r)$ by Lemma~\ref{lem:crossing}.
For $a=13/9$ we have $u=2/3$ and $(1-u)/(1+u)=1/5$, so $b=26/25$. For $a=3/2$ we have $u=1/\sqrt2$ and $(1-u)/(1+u)=(\sqrt2-1)/(\sqrt2+1)=(\sqrt2-1)^2=3-2\sqrt2$, whose square is $17-12\sqrt2$, so $b=18-12\sqrt2$.
\end{proof}

\begin{exa}[Tight instances for $a=13/9$]\label{ex:tight}
Let $a=13/9$, so $\alpha=9/13$.
\emph{Block $(1,1)$:} $p=(1,3/5)$ and $W=(1,6/5)$, so $r=3/5=1/(1+2/3)$ and $w(x)=1+(1-x)/2$. For $x^*$ slightly below $9/13$, we have $p(1)=1>a\,x^*$, so the task is rounded down to two processors with work $6/5$, while $w(x^*)\to w(9/13)=15/13$; the ratio tends to $(6/5)/(15/13)=26/25$.
\emph{Block $(2,1)$:} $p=(1,1,2/3)$ and $W=(1,2,2)$, with non-dominated allotments $1$ and $3$ and $r=2/3$; here $w(x)=1+3(1-x)$. For $x^*\to(9/13)^-$ the task is rounded down to three processors with work $2$, while $w(x^*)\to25/13$; the ratio tends to $26/25$.
Both blocks attain $26/25$ simultaneously; for the block $(2,1)$ the supremum is attained at the left end $r=2/3$ of the feasible interval, where $W(3)=W(2)$, in accordance with Proposition~\ref{prop:21}(b). For $a<13/9$ the block $(2,1)$ becomes worse than $(1,1)$ by Lemma~\ref{lem:crossing}; this is why $13/9$ is the threshold.
\end{exa}

A \emph{per-task rounding rule} chooses for every task an allotment $l'\in\{1,\dots,m\}$ as a function of $m$, the processing times $p(1),\dots,p(m)$ of the task and its value $x^*\in[p(m),p(1)]$; speedup-aware rounding and the threshold rules of~\cite{Jansen06,Jansen11,Chen13} are of this kind.

\begin{prop}[Optimality among per-task roundings]\label{prop:optimal}
Let $a>1$. Every rounding rule that guarantees $p(l')\le a\,x^*$ for all tasks satisfying \Aone, \Atwo, \Athree\ and for all $x^*\in[p(m),p(1)]$ has worst-case work factor $\sup W(l')/w(x^*)\ge b(a)$.
\end{prop}
\begin{proof}
By Lemma~\ref{lem:F-s}, $b(a)-1=\sup F(q,1;r)$, so it suffices to consider $s=1$. Fix a feasible triple $(q,1,r)$ and consider $m=q+1$ and the task with $p=(1,\dots,1,r)$, where the first $q$ entries equal $1$. It satisfies \Aone; its works are $W=(1,2,\dots,q,(q+1)r)$ with $(q+1)r\ge q$, so \Atwo\ holds; and its envelope has a single segment, so \Athree\ holds. Take $x^*\in(r,\alpha)$. The only allotment with $p(l)\le a\,x^*<1$ is $l=q+1$, whose work is $(q+1)r$. Hence the rule has work factor at least $(q+1)r/w(x^*)$, which tends to $1+F(q,1;r)$ as $x^*\to\alpha^-$ by Remark~\ref{rem:worst}. Taking the supremum over all feasible $(q,1,r)$ gives $b(a)$.
\end{proof}

Thus speedup-aware rounding is an optimal per-task rounding for every value of $a$. The instances of the proof use $m=q+1$, so optimality is meant over all $m$; for a fixed $m$ only blocks with $q\le m-1$ exist. Further improvements must come from the analysis of list scheduling or from a stronger LP (Section~\ref{sec:limits}).

\begin{rem}[A variant]\label{rem:variant}
The rule that chooses the smallest allotment $l$ with $p(l)\le a\,x^*$ may round up across several blocks. Its choice is at most the allotment chosen by speedup-aware rounding, which also satisfies the time bound, so by \Atwo\ its work is not larger. It is therefore an $(a,b(a))$-rounding as well (Definition~\ref{def:ab}), and by Proposition~\ref{prop:optimal} it has the same worst case; all results below about speedup-aware rounding hold for this variant as well.
\end{rem}

\section{List Scheduling and the Heavy Path}\label{sec:list}
This section uses only the facts that every final allotment is at most $\mu$ and that the schedule is produced by \LIST\ or by the variant described in Lemma~\ref{lem:es-list}. It applies to every rounding rule.

\begin{defs}[Time slots]\label{def:slots}
For $t\in[0,\Cmax)$ let $\busy(t)=\sum\{l_j: S_j\le t<C_j\}$ be the number of busy processors at time $t$. Let
\[
T_1=\{t:\busy(t)\le\mu-1\},\qquad T_2=\{t:\mu\le\busy(t)\le m-\mu\},\qquad T_3=\{t:\busy(t)\ge m-\mu+1\}.
\]
Since $\busy$ is a step function that changes only at start and completion times, each $T_i$ is a finite union of intervals; we write $|T_i|$ for its total length.
\end{defs}

\begin{lem}[Partition]\label{lem:partition}
If $1\le\mu\le(m+1)/2$, then $T_1,T_2,T_3$ partition $[0,\Cmax)$ and $\Cmax=|T_1|+|T_2|+|T_3|$. For every $a>1$ and every $m\ge1$, the value $\mu=\lceil\beta(a)m\rceil$ satisfies $1\le\mu\le(m+1)/2$.
\end{lem}
\begin{proof}
If $\mu\le(m+1)/2$, then $\mu-1<m-\mu+1$, so the three conditions split the integers $\{0,\dots,m\}$ into three consecutive ranges (the middle one is empty if $\mu>m-\mu$).
For the second statement, the denominator of $\beta(a)=2/(a+2+\sqrt{a^2+4})$ increases with $a$, so $a>1$ implies $\beta(a)<\beta(1)=2/(3+\sqrt5)=(3-\sqrt5)/2<0.382$. As $\beta(a)m>0$, we have $\mu\ge1$. For $m\ge5$, $\mu<\beta m+1$, and $\beta m+1\le(m+1)/2$ is equivalent to $m(1-2\beta)\ge1$; since $1-2\beta>1-(3-\sqrt5)=\sqrt5-2>0.236$, this holds for $m\ge5$. For $m\le4$: if $m=1$, then $\mu=1=(m+1)/2$; if $m=2$, then $2\beta<0.764$ and $\mu=1$; if $m=3$, then $3\beta<1.146$ and $\mu\le2$; if $m=4$, then $4\beta<1.528$ and $\mu\le2$. In all cases $\mu\le(m+1)/2$.
\end{proof}

\begin{lem}[No unforced idleness]\label{lem:no-idle}
Consider the schedule produced by \LIST. If a task $J$ is ready at time $t$, that is, all its predecessors have completed by time $t$, but $S_J>t$, then $\busy(t)\ge m-l_J+1$.
\end{lem}
\begin{proof}
Let $t_0$ be the last decision time with $t_0\le t$; it exists because $0$ is a decision time. The task $J$ becomes ready when its last predecessor completes (at time $0$ if it has none); this is a decision time not later than $t$, hence not later than $t_0$, so $J$ is ready at $t_0$. As $S_J>t\ge t_0$, the scan at $t_0$ examined $J$ without starting it, so fewer than $l_J$ processors were idle when $J$ was examined; the rest of the scan only starts tasks, so at the end of the scan at most $l_J-1$ processors are idle. No decision time lies in $(t_0,t]$, so no task completes there, and tasks start only at decision times. Hence $\busy(t)$ equals the number of busy processors at the end of the scan at $t_0$, which is at least $m-l_J+1$.
\end{proof}

The algorithms of~\cite{Jansen06,Chen13} use another form of list scheduling, which we call \emph{earliest-start list scheduling}: repeatedly, among the tasks whose predecessors have all been placed, compute for each task the earliest time at which it can start in the current partial schedule (after its predecessors complete and with $l_j$ processors available during its whole execution, possibly in an earlier gap), and place a task with the smallest such time.

\begin{lem}\label{lem:es-list}
Lemma~\ref{lem:no-idle} also holds for earliest-start list scheduling.
\end{lem}
\begin{proof}
(a) \emph{Tasks are placed in non-decreasing order of start times.} Let $J$ be placed at some step with start time $S_J$ and let $J'$ be the task placed at the next step. If $J'$ was ready when $J$ was chosen, its earliest start time at that moment was at least $S_J$; placing $J$ only removes available capacity, so the earliest start time of $J'$ cannot decrease, and $S_{J'}\ge S_J$. Otherwise $J'$ became ready only after $J$ was placed, so $J$ is a predecessor of $J'$ and $S_{J'}\ge C_J>S_J$.

(b) Let $J$ be ready at time $t$ (all its predecessors complete by $t$) with $S_J>t$. Call a step \emph{bad} if, at that step, $J$ has all its predecessors placed but is not yet placed, its earliest start time $E$ in the current partial schedule exceeds $t$, and the current partial schedule has at most $m-l_J$ busy processors at time $t$. Suppose some step is bad. Since $J$ cannot start at $t$, some instant of $[t,t+p_J(l_J))$ has more than $m-l_J$ busy processors; since $J$ can start at $E$, every instant of $[E,E+p_J(l_J))$ has at most $m-l_J$ busy processors. As $t<E$, the overloaded instant lies in $(t,E)$, and since the load at $t$ is at most $m-l_J$, some already placed task $K$ starts in $(t,E)$. Consider the step at which $K$ was placed, which comes before the current step. All predecessors of $J$ complete by $t<S_K$, so they start before $S_K$ and, by (a), were placed before $K$; hence $J$ had all its predecessors placed at that step and was not yet placed. As $K$ was chosen, the earliest start time of $J$ at that step was at least $S_K>t$. Fewer tasks had been placed, so the load at $t$ was at most the current one. Hence the step at which $K$ was placed is also bad. Repeating the argument yields an infinite sequence of earlier and earlier bad steps, which is impossible. Hence no step is bad.

(c) Apply (b) to the step at which $J$ is placed: there $E=S_J>t$ and $J$ has all predecessors placed, so the partial schedule must have more than $m-l_J$ busy processors at time $t$. By (a), every task placed later starts at or after $S_J>t$, so it does not run at time $t$. Hence the final schedule has $\busy(t)\ge m-l_J+1$.
\end{proof}

\begin{lem}[Heavy path]\label{lem:heavy}
If all allotments satisfy $l_j\le\mu$, then $G$ contains a path $P_1\prec P_2\prec\dots\prec P_k$ such that every time in $T_1\cup T_2$ lies in one of the intervals $[S(P_i),C(P_i))$, and these intervals are pairwise disjoint.
\end{lem}
\begin{proof}
\emph{Construction.} Let $P_k$ be a task that completes last, so $C(P_k)=\Cmax$. Given $P_i$, stop if it has no predecessor; otherwise let $P_{i-1}$ be a predecessor of $P_i$ that completes last. Each step follows an arc backwards and $G$ is acyclic, so the procedure stops; we renumber the tasks as $P_1\prec\dots\prec P_k$.
\emph{Disjointness.} $P_{i-1}$ is a predecessor of $P_i$, so $S(P_i)\ge C(P_{i-1})$, and the intervals are ordered and disjoint.
\emph{Gaps belong to $T_3$.} Let $i\ge2$ and $t\in[C(P_{i-1}),S(P_i))$. All predecessors of $P_i$ complete by $C(P_{i-1})$, so $P_i$ is ready at $t$ but has not started. By Lemma~\ref{lem:no-idle} (or Lemma~\ref{lem:es-list}), $\busy(t)\ge m-l(P_i)+1\ge m-\mu+1$, so $t\in T_3$. Likewise, $P_1$ has no predecessor and is ready during $[0,S(P_1))$, so this interval is contained in $T_3$.
\emph{Covering.} $[0,\Cmax)$ is the union of $[0,S(P_1))$, the intervals $[S(P_i),C(P_i))$ and the gaps $[C(P_{i-1}),S(P_i))$, because $C(P_k)=\Cmax$. Hence every time outside $T_3$ lies in some $[S(P_i),C(P_i))$.
\end{proof}

\section{A Bound for Every Number of Processors}\label{sec:general}
We now combine the per-task properties with the heavy path. To treat speedup-aware rounding and the rounding of~\cite{Chen13} together, we isolate the properties that the analysis needs.

\begin{defs}[$(a,b)$-rounding]\label{def:ab}
Let $a>1$ and $b\ge1$. A rounding rule is an \emph{$(a,b)$-rounding} for a class of instances if, for every task of every instance in the class, the chosen allotment $l'$ satisfies
\begin{itemize}
\item[(R1)] $p(l')\le a\,x^*$,
\item[(R2)] $W(l')\le b\,w(x^*)$,
\item[(R3)] $W(l')\le m\,x^*$.
\end{itemize}
\end{defs}

By Lemmas~\ref{lem:rounding} and~\ref{lem:F-sign}, speedup-aware rounding with parameter $a$ is an $(a,b(a))$-rounding under \Aone, \Atwo\ and \Athree. The proof of Corollary~\ref{coro:capped} uses only (R2), (R3) and \Atwo, so it holds for every $(a,b)$-rounding:

\begin{lem}[Capped tasks]\label{lem:capped-general}
Under \Atwo, if an $(a,b)$-rounding chooses $l'>\mu$, then $p(\mu)\le(m/\mu)\,x^*$ and $W(\mu)\le b\,w(x^*)$.
\end{lem}

In the rest of this section we assume \Aone, \Atwo\ and \Athree, an $(a,b)$-rounding, the cap $l_j=\min\{l'_j,\mu\}$ with $1\le\mu\le(m+1)/2$, and \LIST\ or earliest-start list scheduling.

\begin{lem}[Path inequality]\label{lem:H}
\[
\Hin\qquad \frac1a|T_1|+\min\Big\{\frac1a,\frac{\mu}{m}\Big\}|T_2|\le C^*.
\]
\end{lem}
\begin{proof}
Let $c=\min\{1/a,\mu/m\}$ and let $P$ be the path of Lemma~\ref{lem:heavy}. For a task $J$ on $P$ let $I_J=[S_J,C_J)$ and $\tau_i(J)=|I_J\cap T_i|$ for $i=1,2,3$; by Lemma~\ref{lem:partition}, $\tau_1(J)+\tau_2(J)+\tau_3(J)=p_J(l_J)$.
If $J$ is not capped ($l_J=l_J'$), then by (R1) $x_J^*\ge p_J(l_J)/a\ge\frac1a\tau_1(J)+\frac1a\tau_2(J)\ge\frac1a\tau_1(J)+c\,\tau_2(J)$.
If $J$ is capped ($l_J=\mu$), then $J$ itself occupies $\mu$ processors while it runs, so $\busy\ge\mu$ on $I_J$ and $\tau_1(J)=0$. By Lemma~\ref{lem:capped-general}, $x_J^*\ge(\mu/m)p_J(\mu)\ge(\mu/m)\tau_2(J)\ge c\,\tau_2(J)=\frac1a\tau_1(J)+c\,\tau_2(J)$.
Summing over the tasks of $P$: by Lemma~\ref{lem:heavy}, the intervals $I_J$ are pairwise disjoint and cover $T_1\cup T_2$, so $\sum_J\tau_1(J)=|T_1|$ and $\sum_J\tau_2(J)=|T_2|$. By Lemma~\ref{lem:paths}, $C^*\ge\sum_{J\in P}x_J^*\ge\frac1a|T_1|+c|T_2|$.
\end{proof}

\begin{lem}[Work inequality]\label{lem:W}
\[
\Win\qquad |T_1|+\mu|T_2|+(m-\mu+1)|T_3|\le b\,m\,C^*.
\]
\end{lem}
\begin{proof}
\emph{Lower bound.} The integral of $\busy(t)$ over $[0,\Cmax)$ equals $\sum_jl_jp_j(l_j)=\sum_jW_j(l_j)$. On $T_1$, $\busy\ge1$, because by Lemma~\ref{lem:heavy} some task of the heavy path is running; on $T_2$, $\busy\ge\mu$, and on $T_3$, $\busy\ge m-\mu+1$, by definition. Hence the left-hand side is at most $\sum_jW_j(l_j)$.
\emph{Upper bound.} If $j$ is not capped, $W_j(l_j)=W_j(l_j')\le b\,w_j(x_j^*)$ by (R2); if $j$ is capped, $W_j(\mu)\le b\,w_j(x_j^*)$ by Lemma~\ref{lem:capped-general}. By Lemma~\ref{lem:paths}, $w_j(x_j^*)\le w_j^*$ and $\sum_jw_j^*\le m\,C^*$. Hence $\sum_jW_j(l_j)\le b\,m\,C^*$.
\end{proof}

The inequality \Win\ counts only one busy processor during $T_1$; this is where the framework loses most (Section~\ref{sec:limits}).

\begin{lem}[The inner linear program]\label{lem:inner}
Let $y_i=|T_i|/C^*$ for $i=1,2,3$, $c=\min\{1/a,\mu/m\}$ and $\nu =m-\mu+1$. Then
\[
\frac{\Cmax}{C^*}=y_1+y_2+y_3\le\max\{A(\mu),B(\mu)\},\qquad A(\mu)=a+\frac{bm-a}{\nu },\quad B(\mu)=\frac{m}{\mu}+\frac{(b-1)m}{\nu }.
\]
\end{lem}
\begin{proof}
Dividing \Hin\ and \Win\ by $C^*>0$ gives $y\ge0$, $\frac1ay_1+c\,y_2\le1$ and $y_1+\mu y_2+\nu y_3\le b\,m$; the equality follows from Lemma~\ref{lem:partition}. We bound $y_1+y_2+y_3$ by weak duality, with multipliers $\lambda_W=1/\nu $ for the second inequality and
\[
\lambda_H=\max\Big\{a\Big(1-\frac1\nu \Big),\ \frac{1-\mu/\nu }{c}\Big\}
\]
for the first. Since $\mu\le(m+1)/2$, we have $\nu \ge\mu\ge1$, so both terms are non-negative. The coefficient of $y_1$ in the combination is $\lambda_H/a+\lambda_W\ge(1-1/\nu )+1/\nu =1$; the coefficient of $y_2$ is $\lambda_Hc+\lambda_W\mu\ge(1-\mu/\nu )+\mu/\nu =1$; the coefficient of $y_3$ is $\lambda_W\nu=1$. Since $y\ge0$,
\[
y_1+y_2+y_3\le\lambda_H\Big(\frac1ay_1+c\,y_2\Big)+\lambda_W\big(y_1+\mu y_2+\nu y_3\big)\le\lambda_H+\frac{bm}{\nu }.
\]
If $\lambda_H=a(1-1/\nu )$, the right-hand side is $a-a/\nu +bm/\nu =A(\mu)$. If $\lambda_H=(1-\mu/\nu )/c$ and $c=\mu/m$, it is $(m/\mu)(1-\mu/\nu )+bm/\nu =m/\mu-m/\nu +bm/\nu =B(\mu)$. If $\lambda_H=(1-\mu/\nu )/c$ and $c=1/a$, it is $a(1-\mu/\nu )+bm/\nu \le a(1-1/\nu )+bm/\nu =A(\mu)$, because $\mu\ge1$. In all cases the bound is at most $\max\{A(\mu),B(\mu)\}$.
\end{proof}

\begin{lem}[The functions $\beta$ and $h$]\label{lem:beta}
For $a>0$ let $\beta(a)=2/(a+2+\sqrt{a^2+4})$ and $h(a)=(2-a+\sqrt{a^2+4})/2$.
\begin{itemize}
\item[(a)] $\beta(a)$ is the root in $(0,1/2)$ of $a\beta^2-(a+2)\beta+1=0$, and it satisfies $a+1/(1-\beta)=1/\beta$.
\item[(b)] $1/(1-\beta(a))=h(a)$.
\item[(c)] $1/\beta(a)=a+h(a)=(a+2+\sqrt{a^2+4})/2$.
\end{itemize}
\end{lem}
\begin{proof}
(a) The discriminant is $(a+2)^2-4a=a^2+4$, so the roots are $((a+2)\pm\sqrt{a^2+4})/(2a)$. Rationalizing the smaller root, $((a+2)-\sqrt{a^2+4})/(2a)=4a/\big(2a(a+2+\sqrt{a^2+4})\big)=\beta(a)$. Clearly $\beta>0$, and $\beta<1/2$ is equivalent to $\sqrt{a^2+4}>2-a$, which holds trivially for $a\ge2$ and, after squaring, reduces to $4a>0$ for $a<2$. Multiplying $a+1/(1-\beta)=1/\beta$ by $\beta(1-\beta)>0$ gives $a\beta(1-\beta)+\beta=1-\beta$, which is the quadratic equation.
(b) $1-\beta=(a+\sqrt{a^2+4})/(a+2+\sqrt{a^2+4})$, so $1/(1-\beta)=(a+2+\sqrt{a^2+4})/(a+\sqrt{a^2+4})$. Multiplying numerator and denominator by $\sqrt{a^2+4}-a$ turns the denominator into $4$ and the numerator into $2\sqrt{a^2+4}-2a+4$, which gives $(2-a+\sqrt{a^2+4})/2$.
(c) By (a) and (b), $1/\beta=a+1/(1-\beta)=a+h(a)$.
\end{proof}

\begin{thm}[General bound]\label{thm:general}
Assume \Aone, \Atwo\ and \Athree, and let $a>1$ and $b\ge1$. For every $m\ge1$, every $(a,b)$-rounding followed by the cap $\mu=\lceil\beta(a)m\rceil$ and by \LIST\ (or earliest-start list scheduling) produces a schedule with
\[
\Cmax\le\big(a+b\,h(a)\big)\,C^*\le\big(a+b\,h(a)\big)\,\OPT.
\]
\end{thm}
\begin{proof}
Write $\beta=\beta(a)$ and $\nu =m-\mu+1$. By Lemma~\ref{lem:partition}, $1\le\mu\le(m+1)/2$, so Lemma~\ref{lem:inner} applies.
By definition of the ceiling, $\beta m\le\mu<\beta m+1$, so $\nu >m-\beta m=(1-\beta)m>0$.
Then $A(\mu)=a+(bm-a)/\nu <a+bm/\nu <a+bm/((1-\beta)m)=a+b/(1-\beta)$, using $a>0$ and $\nu >(1-\beta)m$.
Further, $B(\mu)=m/\mu+(b-1)m/\nu \le1/\beta+(b-1)/(1-\beta)$, using $\mu\ge\beta m$, $b\ge1$ and $\nu >(1-\beta)m$.
By Lemma~\ref{lem:beta}(a), $1/\beta=a+1/(1-\beta)$, so $1/\beta+(b-1)/(1-\beta)=a+b/(1-\beta)$.
Lemma~\ref{lem:inner} and Lemma~\ref{lem:beta}(b) give $\Cmax/C^*\le\max\{A(\mu),B(\mu)\}\le a+b/(1-\beta)=a+b\,h(a)$, and Lemma~\ref{lem:relax} gives $C^*\le\OPT$.
\end{proof}

\begin{thm}[Speedup-aware rounding]\label{thm:main}
Assume \Aone, \Atwo\ and \Athree. For every $a>1$ and every $m\ge1$, algorithm $\SAR(a,\lceil\beta(a)m\rceil)$ satisfies $\Cmax\le R(a)\,\OPT$ with $R(a)=a+b(a)\,h(a)$.
\end{thm}
\begin{proof}
By Lemmas~\ref{lem:rounding} and~\ref{lem:F-sign}, speedup-aware rounding is an $(a,b(a))$-rounding with $b(a)\ge1$, and $b(a)<\infty$ by Lemma~\ref{lem:small-q}. Apply Theorem~\ref{thm:general}.
\end{proof}

\begin{rem}[Why $\beta(a)$]\label{rem:why-beta}
As $m\to\infty$ with $\mu/m\to\beta$, $A\to a+b/(1-\beta)$, which increases with $\beta$, and $B\to1/\beta+(b-1)/(1-\beta)$. The derivative of the latter, $-1/\beta^2+(b-1)/(1-\beta)^2$, is negative for $\beta<1/(1+\sqrt{b-1})$, which is at least $1/2>\beta(a)$ whenever $b\le2$ (and exceeds $0.75$ when $b\le1.1$). The two limits are equal exactly when $a+1/(1-\beta)=1/\beta$, a condition that does not involve $b$ and defines $\beta(a)$; to the left of this point $B$ is larger and to the right $A$ is larger, so $\beta(a)$ minimizes the limit of $\max\{A,B\}$. By Lemma~\ref{lem:beta}(c), $R(a)=1/\beta(a)+(b(a)-1)h(a)$: the first term is the value for a perfect rounding ($b=1$) and the second term is the price of rounding.
\end{rem}

\begin{rem}[Choosing $\mu$ for a given $m$]\label{rem:mu}
Lemma~\ref{lem:inner} holds for every integer $\mu$ with $1\le\mu\le(m+1)/2$. An implementation can therefore pick the $\mu$ that minimizes $\max\{A(\mu),B(\mu)\}$, in $O(m)$ time; by Theorem~\ref{thm:general}, the resulting guarantee is never worse than $a+b\,h(a)$. In the same way one may, for a given $m$, use the parameter $a$ whose guarantee is smaller; for instance, $a\ge2$ (rounding up only, $b=1$) is better for $m=2$. Table~\ref{tab:finite} in Section~\ref{sec:numerics} lists such guarantees.
\end{rem}

\section{Choosing the Parameter \texorpdfstring{$a$}{a}}\label{sec:opt-a}
We minimize $R(a)=a+b(a)h(a)$. On $[13/9,2)$, Theorem~\ref{thm:closed} gives $b(a)$ explicitly; for $a<13/9$ we use the lower bound of Proposition~\ref{prop:21}(a), and for $a\ge2$ we have $b(a)=1$.

\begin{lem}[Derivatives]\label{lem:deriv}
\textup{(a)} $h'(a)=(a/\sqrt{a^2+4}-1)/2<0$ and $h''(a)=2/(a^2+4)^{3/2}>0$; hence $h$ is decreasing and strictly convex, and $|h'(a)|=(1-a/\sqrt{a^2+4})/2$ decreases in $a$.
\textup{(b)} $1/\beta(a)=(a+2+\sqrt{a^2+4})/2$ has derivative $(1+a/\sqrt{a^2+4})/2>0$.
\textup{(c)} On $(1,2)$, $g(a)=((1-u)/(1+u))^2$ with $u=\sqrt{a-1}$ satisfies $g'(a)=-2(1-u)/(u(1+u)^3)<0$, and $|g'(a)|$ decreases in $a$; hence $g$ is decreasing and convex.
\end{lem}
\begin{proof}
(a) $h=(2-a+(a^2+4)^{1/2})/2$, so $h'=(-1+a(a^2+4)^{-1/2})/2<0$ because $a<\sqrt{a^2+4}$, and $h''=\big((a^2+4)^{-1/2}-a^2(a^2+4)^{-3/2}\big)/2=2/(a^2+4)^{3/2}$. The function $a/\sqrt{a^2+4}$ has derivative $4/(a^2+4)^{3/2}>0$, so $|h'|$ decreases.
(b) Direct differentiation.
(c) $\frac{d}{du}\frac{1-u}{1+u}=-\frac{2}{(1+u)^2}$, so $\frac{dg}{du}=-\frac{4(1-u)}{(1+u)^3}$; with $\frac{du}{da}=\frac1{2u}$ we get $g'(a)$. The numerator $2(1-u)$ decreases and the denominator $u(1+u)^3$ increases with $u$, which increases with $a$; hence $|g'|$ decreases, $g'$ increases, and $g$ is convex.
\end{proof}

\begin{thm}[The best parameter under \Athree]\label{thm:opt-a}
On $a>1$, $R(a)$ attains its minimum at $a=13/9$, where
\[
R(13/9)=\frac{13}{9}+\frac{26}{25}\cdot\frac{5+\sqrt{493}}{18}=\frac{13(30+\sqrt{493})}{225}\approx3.016208.
\]
Consequently, under \Aone, \Atwo\ and \Athree, algorithm $\SAR(13/9,\lceil18m/(31+\sqrt{493})\rceil)$ is a $13(30+\sqrt{493})/225$-approximation for every $m$.
\end{thm}
\begin{proof}
For $a=13/9$ we have $2-a=5/9$ and $a^2+4=493/81$, so $h(13/9)=(5+\sqrt{493})/18$ and $\beta(13/9)=18/(31+\sqrt{493})$. From $22.2036^2<493<22.2037^2$ we obtain $h(13/9)\in(1.51131,1.51132)$ and $|h'(13/9)|=(1-13/\sqrt{493})/2\in(0.20725,0.20726)$. Moreover $R(13/9)=13/9+\frac{26}{25}\cdot\frac{5+\sqrt{493}}{18}=\frac{325+13(5+\sqrt{493})}{225}=\frac{390+13\sqrt{493}}{225}$. We compare three ranges.

\emph{$a\ge2$.} By Lemma~\ref{lem:small-q}, $b(a)=1$, so $R(a)=a+h(a)=1/\beta(a)\ge1/\beta(2)=(4+\sqrt8)/2=2+\sqrt2\approx3.414>R(13/9)$, using Lemmas~\ref{lem:beta}(c) and~\ref{lem:deriv}(b).

\emph{$13/9\le a<2$.} By Theorem~\ref{thm:closed}, $R(a)=a+(1+g(a))h(a)$ is differentiable with $R'(a)=1+g'(a)h(a)+(1+g(a))h'(a)$. By Lemma~\ref{lem:deriv}: $g'<0$ and $|g'(a)|\le|g'(13/9)|=2\cdot\frac13/\big(\frac23(\frac53)^3\big)=27/125$; $h(a)\le h(13/9)<1.51132$; $1+g(a)\le1+g(13/9)=26/25$; and $|h'(a)|\le|h'(13/9)|<0.20726$. Hence $R'(a)\ge1-\frac{27}{125}\cdot1.51132-\frac{26}{25}\cdot0.20726>1-0.32645-0.21556>0.45$, so $R$ is increasing on this range and $R(a)\ge R(13/9)$.

\emph{$1<a\le13/9$.} By Proposition~\ref{prop:21}(a), $b(a)\ge2a/(4a-3)$, so $R(a)\ge\varphi(a):=a+\frac{2a}{4a-3}h(a)$. Since $\frac{d}{da}\frac{2a}{4a-3}=-\frac{6}{(4a-3)^2}$, we have $\varphi'(a)=1-\frac{6h(a)}{(4a-3)^2}+\frac{2a}{4a-3}h'(a)$. On this range $4a-3\in(1,25/9]$ and $h(a)\ge h(13/9)>1.51131$, so $\frac{6h(a)}{(4a-3)^2}\ge6\cdot1.51131\cdot\frac{81}{625}>1.1751$; moreover $\frac{2a}{4a-3}>0$ and $h'<0$. Hence $\varphi'(a)<1-1.1751<0$, $\varphi$ is decreasing, and $R(a)\ge\varphi(a)\ge\varphi(13/9)=13/9+\frac{26}{25}h(13/9)=R(13/9)$.

Combining the three ranges proves the first statement; the second follows from Theorem~\ref{thm:main}.
\end{proof}

\begin{coro}[Simple parameters]\label{coro:three-half}
Under \Aone, \Atwo\ and \Athree, algorithm $\SAR(3/2,\lceil m/3\rceil)$ satisfies $\Cmax\le(57/2-18\sqrt2)\,\OPT$ for every $m$; note that $57/2-18\sqrt2\approx3.044156$.
\end{coro}
\begin{proof}
By Theorem~\ref{thm:closed}, $b(3/2)=18-12\sqrt2$. Further $\beta(3/2)=2/(3/2+2+5/2)=1/3$ and $h(3/2)=1/(1-1/3)=3/2$ by Lemma~\ref{lem:beta}(b). Theorem~\ref{thm:main} gives the ratio $3/2+\frac32(18-12\sqrt2)=57/2-18\sqrt2$.
\end{proof}

\begin{rem}[Earlier rules in the same framework]\label{rem:old-rules}
If the work factor $b(a)$ is replaced by $a/(a-1)$, the trade-off of the threshold rule of~\cite{Jansen06} with $\rho=1/a$, then $\min_a\{a+\frac{a}{a-1}h(a)\}\approx4.730577$, attained at $a\approx2.3202$, that is, $\rho=1/a\approx0.4310$; for $\rho=0.43$, the value used in~\cite{Jansen06}, the expression equals $4.730598$, the ratio reported there. Rounding up only ($a=2$, $b=1$) gives $1/\beta(2)=2+\sqrt2$, the ratio of~\cite{Chen13}. The whole improvement of this paper under \Athree\ therefore comes from the smaller work factor $b(a)$.
\end{rem}

\section{Strictly Decreasing Processing Times and Concave Speedup}\label{sec:strict}
Under \Aonec, every block consists of a single allotment, and the work factor equals $1+g(a)$ for all $a\in(1,2)$.

\begin{thm}\label{thm:strict}
Assume \Aonec, \Atwo\ and \Athree, and let $1<a<2$. Then speedup-aware rounding is an $(a,b_s(a))$-rounding with
\[
b_s(a)=1+g(a)=1+\Big(\frac{1-\sqrt{a-1}}{1+\sqrt{a-1}}\Big)^2,
\]
and $\SAR(a,\lceil\beta(a)m\rceil)$ satisfies $\Cmax\le R_s(a)\,\OPT$ with $R_s(a)=a+(1+g(a))h(a)$ for every $m$.
\end{thm}
\begin{proof}
Fix a task. Under \Aonec\ the non-dominated allotments are exactly $1,2,\dots,l^\circ$: for $l\le l^\circ$, $p(l)<p(l-1)$, and for $l>l^\circ$, $p(l)=p(l-1)$. Hence consecutive non-dominated allotments are $l$ and $l+1$, and every block $[s..q]$ has $q=s$. The allotments $l^\circ,\dots,m$ do not form a block, since no faster allotment follows; and as $x^*\ge p(m)=p(l^\circ)$ by~\eqref{lp:range}, the value $x^*$ never lies strictly inside this tail.
In the proof of Lemma~\ref{lem:rounding}, $b(a)$ is used only in Case~2, and all triples $(q,s,r)$ arising there now have $s=q$. Hence Lemma~\ref{lem:rounding}(ii) holds with $b(a)$ replaced by $1+\sup\{F(q,q;r):(q,q,r)\text{ feasible}\}$. By Lemma~\ref{lem:singleton}, $\sup_rF(q,q;r)\le\sup_rF(1,1;r)$, which equals $g(a)$ by Proposition~\ref{prop:11}. Properties (R1) and (R3) are unchanged. Theorem~\ref{thm:general} with $b=1+g(a)$ completes the proof.
\end{proof}

\begin{prop}[The best parameter under \Aonec]\label{prop:strict-opt}
The function $R_s$ is strictly convex on $(1,2)$ and has a unique minimizer $a_s$; numerically $a_s\approx1.284814$ and $R_s(a_s)\approx2.973912$. Without solving for $a_s$, the choice $a=9/7$ gives $b_s(9/7)=(162-36\sqrt{14})/25$, $h(9/7)=(5+\sqrt{277})/14$ and
\[
R_s(9/7)=\frac97+\frac{(162-36\sqrt{14})(5+\sqrt{277})}{350}\approx2.973914.
\]
\end{prop}
\begin{proof}
\emph{Convexity.} $R_s=a+h+g\,h$. The term $a$ is affine and $h$ is strictly convex (Lemma~\ref{lem:deriv}). The functions $g$ and $h$ are positive, decreasing and convex, so $(gh)''=g''h+2g'h'+gh''\ge0$, where $g'h'>0$ because both derivatives are negative. Hence $R_s''\ge h''>0$.
\emph{Unique minimizer.} $R_s'=1+h'+g'h+gh'$. As $a\to1^+$, $u\to0$ and $g'(a)=-2(1-u)/(u(1+u)^3)\to-\infty$, so $R_s'\to-\infty$. As $a\to2^-$, $u\to1$, $g\to0$ and $g'\to0$, so $R_s'\to1+h'(2)=1+(2/\sqrt8-1)/2\approx0.854>0$. Since $R_s'$ is strictly increasing, it has exactly one zero $a_s$, which is the minimizer; numerically $a_s\approx1.2848137$.
\emph{The value at $9/7$.} Here $u=\sqrt{2/7}$ and $(1-u)/(1+u)=(\sqrt7-\sqrt2)/(\sqrt7+\sqrt2)=(\sqrt7-\sqrt2)^2/5=(9-2\sqrt{14})/5$, whose square is $(137-36\sqrt{14})/25$; hence $b_s(9/7)=(162-36\sqrt{14})/25\approx1.092013$. Further $a^2+4=277/49$, so $h(9/7)=(5+\sqrt{277})/14$. Substituting into $R_s=a+b_sh$ and using $25\cdot14=350$ gives the formula.
\end{proof}

\begin{prop}[Concave speedup]\label{prop:concave}
For a task, let $S(l)=p(1)/p(l)$ for $l=1,\dots,m$, let $S(0)=0$, and let $\Delta_l=S(l)-S(l-1)$. The speedup $S$ is non-decreasing and concave on $\{0,1,\dots,m\}$, that is, $\Delta_1\ge\Delta_2\ge\dots\ge\Delta_m\ge0$, if and only if the task satisfies \Aonec, \Atwo\ and \Athree. In particular, under \Aonep, a concave speedup is equivalent to \Atwo\ and \Athree, so the setting of~\cite{Jansen11} is \Aonep, \Atwo\ and \Athree, and the ratio $R_s(a_s)\approx2.973912$ of Theorem~\ref{thm:strict} applies to it.
\end{prop}
\begin{proof}
By Fact~\ref{fact:scaling} let $p(1)=1$, so that $p(l)=1/S(l)$, $W(l)=l/S(l)$ and $\Delta_1=1$.
\emph{A formula for the marginal works.} If $\Delta_{l+1}>0$, then
\[
W(l+1)-W(l)=\frac{(l+1)S(l)-l\,S(l+1)}{S(l)S(l+1)},\qquad p(l)-p(l+1)=\frac{\Delta_{l+1}}{S(l)S(l+1)},
\]
so the marginal work between the allotments $l$ and $l+1$ is $\sigma_l=\big((l+1)S(l)-lS(l+1)\big)/\Delta_{l+1}$; substituting $S(l+1)=S(l)+\Delta_{l+1}$ gives $\sigma_l=S(l)/\Delta_{l+1}-l$. If moreover $\Delta_{l+2}>0$, then, since $S(l)/\Delta_{l+1}+1=S(l+1)/\Delta_{l+1}$,
\begin{equation}\label{eq:sigma-diff}
\sigma_{l+1}-\sigma_l=\frac{S(l+1)}{\Delta_{l+2}}-\frac{S(l)}{\Delta_{l+1}}-1=S(l+1)\Big(\frac{1}{\Delta_{l+2}}-\frac{1}{\Delta_{l+1}}\Big),
\end{equation}
and because $S(l+1)>0$, we have $\sigma_{l+1}\ge\sigma_l$ if and only if $\Delta_{l+2}\le\Delta_{l+1}$.

\emph{Concave speedup implies \Aonec, \Atwo, \Athree.} \Aone: $\Delta_l\ge0$, so $S$ is non-decreasing and $p=1/S$ is non-increasing. Final plateau: if $\Delta_{l+1}=0$, that is $p(l+1)=p(l)$, then $0\le\Delta_k\le\Delta_{l+1}=0$ for all $k>l+1$, so $S$ and $p$ are constant from $l$ on; hence $p$ strictly decreases until it stops decreasing, which is \Aonec. \Atwo: $S(l)/l=(\Delta_1+\dots+\Delta_l)/l$ is the average of the first $l$ increments; $\Delta_{l+1}$ is at most each of them, so adding it does not increase the average, $S(l+1)/(l+1)\le S(l)/l$, and $W(l)=l/S(l)$ is non-decreasing. \Athree: the non-dominated allotments are $1,\dots,l^\circ$, and for consecutive marginal works $\sigma_l,\sigma_{l+1}$ with $l+2\le l^\circ$ we have $0<\Delta_{l+2}\le\Delta_{l+1}$, so $\sigma_{l+1}\ge\sigma_l$ by~\eqref{eq:sigma-diff}; hence $w$ is convex by Definition~\ref{def:block}.

\emph{\Aonec, \Atwo, \Athree\ imply a concave speedup.} By \Aonec, $\Delta_{l+1}>0$ for $l<l^\circ$ and $\Delta_{l+1}=0$ for $l\ge l^\circ$. If $l^\circ\ge2$, \Atwo\ gives $W(2)\ge W(1)$, that is $2/S(2)\ge1$, so $\Delta_2=S(2)-1\le1=\Delta_1$. For $1\le l\le l^\circ-2$, \Athree\ gives $\sigma_{l+1}\ge\sigma_l$, hence $\Delta_{l+2}\le\Delta_{l+1}$ by~\eqref{eq:sigma-diff}. Finally, $\Delta_k=0\le\Delta_{k-1}$ for $k>l^\circ$. Hence $\Delta_1\ge\Delta_2\ge\dots\ge\Delta_m\ge0$.
\end{proof}

\begin{rem}[Comparison with~\cite{Jansen11}]\label{rem:jz12}
The work constraints of the LP of~\cite{Jansen11} contain the factor $1/(p_j(l)-p_j(l+1))$ for every $l$, so that LP is undefined as soon as $p_j(l)=p_j(l+1)$ for some $l$; this is why~\cite{Jansen11} effectively requires \Aonep, as observed in~\cite{Chen13}. By Proposition~\ref{prop:concave}, the setting of~\cite{Jansen11} is therefore exactly \Aonep, \Atwo\ and \Athree, and for it Theorem~\ref{thm:strict} improves the ratio $\approx3.2919$ to $\approx2.973912$. The LP $(\mathrm{P})$ uses marginal works only between consecutive non-dominated allotments, whose processing times differ, and treats allotments with equal processing times as one block; the analysis handles blocks through the bound $r\ge q/(q+1)$ of Fact~\ref{fact:ratio}. Hence $(\mathrm{P})$ and Theorem~\ref{thm:strict} do not need \Aonep. Concave speedup functions with a final plateau ($p(l^\circ)=\dots=p(m)$ with $l^\circ<m$) violate \Aonep, so the LP of~\cite{Jansen11} is not defined for them as stated. After the allotments beyond $l^\circ$ are discarded, which does not change $\OPT$ (Lemma~\ref{lem:useless}), every task satisfies \Aonep\ on $\{1,\dots,l^\circ\}$, and the algorithm and analysis of~\cite{Jansen11} apply: after the restriction the allotments of every task form a set $\{1,\dots,l^\circ_j\}$ with $l^\circ_j\le m$, on which their assumptions hold, and their Lemmas~4.1--4.4 use only these properties. The previous ratio for this class is therefore $\approx3.2919$ as well, and Theorem~\ref{thm:strict} improves it to $\approx2.973912$. The class that lies beyond~\cite{Jansen11} is \Aone\ with plateaus inside the range of allotments. Discarding the dominated allotments there leaves allotments $s_1<\dots<s_K$ that are not consecutive, and the work factor of threshold rounding then approaches $1/(1-\rho)$ (Remark~\ref{rem:threshold}); the blocks of Definition~\ref{def:block} handle this case, and Theorem~\ref{thm:opt-a} gives $\approx3.016208$ there.
\end{rem}

\section{The Strictly Decreasing Case of Chen and Chu Revisited}\label{sec:cc13}
For processing times that strictly decrease in the number of processors, Chen and Chu~\cite{Chen13} reported the approximation ratio $2.9549$ for their algorithm. In this section we point out the step of the proof that is not valid, observe that in this case the algorithm coincides with that of Jansen and Zhang~\cite{Jansen11} up to its parameters, and prove the guarantee of the algorithm with parameters chosen by the corrected analysis. The guarantee holds under the weaker condition \Aonec\ as well.

\subsection{The algorithm of Chen and Chu in our notation}\label{sec:cc13-alg}
By Remark~\ref{rem:lp-relation}, the LP of~\cite{Chen13} is $(\mathrm{P})$ under \Athree, up to the constraint $x_j\le C_j$, which we include. The rounding of~\cite{Chen13} is a threshold rule: with a parameter $\rho\in[0,1)$, a task whose value $x^*$ lies strictly inside the segment of a block $[s..q]$ is rounded up to $s$ if $\lambda=(x^*-p(q+1))/(p(s)-p(q+1))\ge\rho$ and down to $q+1$ otherwise; if $x^*$ equals the processing time of a non-dominated allotment, that allotment is used (for $\rho=0$ the rule as printed differs at these points, see Remark~\ref{rem:large-a}; this does not affect the analysis). The allotments are then capped at $\mu$, and earliest-start list scheduling is applied. Under \Aonep, and more generally under \Aonec, every block is a single allotment, so $q=s$ (see the proof of Theorem~\ref{thm:strict}).

\textbf{Coincidence with~\cite{Jansen11}.} Under \Aonep\ and \Athree, the LP~(9) of~\cite{Jansen11} is the epigraph form of the envelope work function, that is, $(\mathrm{P})$ without the constraint $x_j\le C_j$ (Remark~\ref{rem:lp-relation}). Its rounding uses the critical point $\rho\,p(l)+(1-\rho)p(l+1)$ and rounds up exactly when $x^*$ is at least this point, that is, when $\lambda\ge\rho$; and it caps the allotments at $\mu$ and applies earliest-start list scheduling. Hence, for strictly decreasing processing times, the algorithms of~\cite{Chen13} and~\cite{Jansen11} differ only in the choice of $\rho$ and $\mu$ (and possibly in the choice among optimal LP solutions).

\textbf{Final plateaus.} The analysis below uses \Aonep\ only through $q=s$, which holds under \Aonec\ as well, so we state it under \Aonec. Equivalently, the strictly decreasing case already covers \Aonec, because a final plateau can be removed. Consider a task with $l^\circ_j<m$. Its allotments $l^\circ_j+1,\dots,m$ are dominated and never needed: the LP solution satisfies $x_j^*\ge p_j(m)=p_j(l^\circ_j)$, so $x_j^*$ never lies strictly inside the plateau, and the threshold rule chooses only non-dominated allotments. Removing these allotments, as in Lemma~\ref{lem:useless}, changes neither $\OPT$ nor $(\mathrm{P})$, since the lower bound in~\eqref{lp:range} becomes $p_j(l^\circ_j)=p_j(m)$ and~\eqref{lp:work} involves only non-dominated allotments; hence it does not change the output of the algorithm either. Afterwards the processing times of the task strictly decrease on its allotments $\{1,\dots,l^\circ_j\}$, and the analysis uses only that allotments do not exceed $m$. By Proposition~\ref{prop:concave}, the guarantee of Theorem~\ref{thm:cc13} therefore holds for every concave speedup; after the removal, the algorithm of~\cite{Jansen11} is defined for such tasks as well (Remark~\ref{rem:jz12}).

\subsection{The step that fails}\label{sec:cc13-fail}
The analysis of~\cite{Chen13} normalizes $x_i=|T_i|/C^*$ for $i=1,2,3$ and $x_p=P/C^*$, where $P=\sum_jp_j(1)$. Under \Aonep\ it uses three inequalities: the path inequality $\frac{1+\rho}{2}x_1+\min\{\frac{1+\rho}{2},\frac{\mu}{m}\}x_2\le1$; the bound $\Cmax\le P$~\cite[Lemma~4.5]{Chen13}, that is, $x_1+x_2+x_3\le x_p$; and the work bound $W\le W^*+\rho P$~\cite[Lemmas~4.13 and~4.14]{Chen13}, where $W^*$ is the LP work, which together with $W\ge|T_1|+\mu|T_2|+(m-\mu+1)|T_3|$ and $W^*\le mC^*$ gives
\begin{equation}\label{eq:cc13-work}
x_1+\mu x_2+(m-\mu+1)x_3\le m+\rho\,x_p
\end{equation}
\cite[Lemma~4.15]{Chen13}. The proof of~\cite[Lemma~4.16]{Chen13} then eliminates $x_p$ by replacing it with $x_1+x_2+x_3$ in~\eqref{eq:cc13-work}. In the maximization over $(x_1,x_2,x_3,x_p)$, however, $x_p$ appears on the right-hand side of~\eqref{eq:cc13-work} with a positive coefficient, so a larger $x_p$ only relaxes the constraint. The replacement would require $x_p\le x_1+x_2+x_3$, which is the opposite of~\cite[Lemma~4.5]{Chen13}. The only upper bound available is $P\le W^*\le mC^*$, that is, $x_p\le m$. (In the general case of~\cite{Chen13} the corresponding work bound reads $W\le(W^*-\rho P)/(1-\rho)$; its right-hand side decreases in $P$, and the same replacement is valid there.)

With the parameters $m=8$, $\mu=4$ and $\rho=0.5921$ of~\cite[Table~3]{Chen13}, the program with $x_p\le m$ has the value $3.5523$, attained at $x_1=1/0.79605$, $x_2=0$, $x_3\approx2.2961$ and $x_p=8$, whereas~\cite{Chen13} reports $2.9549$ for this $m$. The bound $2.9549$ therefore does not follow from the lemmas of~\cite{Chen13}. This concerns the proof only; it does not show that the algorithm performs worse on actual instances.

\subsection{A corrected analysis}
The repair is to bound the work task by task instead of through $P$. The two per-task factors below are those of~\cite[Lemma~4.2]{Jansen11}; we include short proofs for completeness and because Lemma~\ref{lem:cc13-time} also covers blocks.

\begin{lem}[Time factor of threshold rounding]\label{lem:cc13-time}
Assume \Aone\ and \Atwo, and let $0\le\rho<1$. The threshold rule chooses an allotment $l'$ with $p(l')\le\frac{2}{1+\rho}\,x^*$.
\end{lem}
\begin{proof}
If $x^*$ is the processing time of a non-dominated allotment, then $p(l')=x^*$. If the task is rounded down, $p(l')=p(q+1)<x^*$. If it is rounded up, normalize $p(s)=1$ and $p(q+1)=r$ (Fact~\ref{fact:scaling}); then $x^*=\lambda+(1-\lambda)r=r+\lambda(1-r)\ge r+\rho(1-r)=\rho+(1-\rho)r$. By Fact~\ref{fact:ratio}, $r\ge q/(q+1)\ge1/2$, so $x^*\ge\rho+(1-\rho)/2=(1+\rho)/2$, that is, $p(s)\le\frac{2}{1+\rho}x^*$.
\end{proof}

\begin{lem}[Work factor of threshold rounding under \Aonec]\label{lem:cc13-work}
Assume \Aonec\ and \Atwo, and let $0\le\rho<1$. The threshold rule chooses an allotment $l'$ with $W(l')\le\frac{2}{2-\rho}\,w(x^*)$ and $W(l')\le m\,x^*$.
\end{lem}
\begin{proof}
If $x^*$ is the processing time of an allotment, then $W(l')=w(x^*)$. If the task is rounded up, $W(l')=W(s)=w(p(s))\le w(x^*)$ by Fact~\ref{fact:monotone}. In both cases $W(l')\le w(x^*)\le m\,x^*$ by Fact~\ref{fact:mx}.
Suppose the task is rounded down. Under \Aonec, $q=s$, so $l'=s+1$, $\lambda<\rho$ and, by~\eqref{eq:envelope}, $w(x^*)=\lambda W(s)+(1-\lambda)W(s+1)$. This expression does not increase with $\lambda$, because $W(s)\le W(s+1)$ by \Atwo; hence $w(x^*)\ge\rho\,W(s)+(1-\rho)W(s+1)$. With $W(s)=s\,p(s)\ge s\,p(s+1)$ we obtain $w(x^*)\ge(s+1-\rho)\,p(s+1)$, so
\[
\frac{W(s+1)}{w(x^*)}\le\frac{s+1}{s+1-\rho}\le\frac{2}{2-\rho},
\]
because $(s+1)/(s+1-\rho)$ decreases in $s\ge1$. Finally $W(s+1)=(s+1)p(s+1)\le m\,p(s+1)<m\,x^*$.
\end{proof}

\begin{rem}[Tightness]\label{rem:cc13-tight}
The factor $2/(2-\rho)$ cannot be improved for the threshold rule. Let $m=2$, $p=(1,1-\varepsilon)$ and let $x^*$ approach $\rho+(1-\rho)(1-\varepsilon)$ from below, so that $\lambda\to\rho^-$ and the task is rounded down. Then $W(2)/w(x^*)\to2(1-\varepsilon)/\big(\rho+2(1-\varepsilon)(1-\rho)\big)$, which tends to $2/(2-\rho)$ as $\varepsilon\to0$. Speedup-aware rounding treats this task differently: the second processor saves almost no time, so $p(1)\le a\,x^*$ and the task is rounded up at no extra work.
\end{rem}

\begin{thm}[The strictly decreasing case of~\cite{Chen13}]\label{thm:cc13}
Assume \Aonec, \Atwo\ and \Athree, let $0\le\rho<1$ and $a_\rho=2/(1+\rho)$. For every $m$, the algorithm of~\cite{Chen13} (equivalently, of~\cite{Jansen11} after final plateaus are removed) with parameter $\rho$, with the cap $\mu=\lceil\beta(a_\rho)m\rceil$ and with the LP $(\mathrm{P})$, satisfies
\[
\Cmax\le\Big(\frac{2}{1+\rho}+\frac{2}{2-\rho}\,h\Big(\frac{2}{1+\rho}\Big)\Big)\OPT.
\]
For $\rho=0.26$ the bound equals $100/63+100(\sqrt{6469}+13)/5481\approx3.291919$, with $\mu=\lceil63m/(113+\sqrt{6469})\rceil$; this cap behaves asymptotically like the cap $(113m-\sqrt{6469m^2-6300m})/100$ of~\cite{Jansen11}. Its infimum over $\rho$ is $\approx3.291913$, attained at $\rho\approx0.26192$.
\end{thm}
\begin{proof}
For $\rho<1$ we have $a_\rho>1$. By Lemmas~\ref{lem:cc13-time} and~\ref{lem:cc13-work}, the threshold rule is an $(a_\rho,2/(2-\rho))$-rounding under \Aonec, \Atwo\ and \Athree, and $2/(2-\rho)\ge1$. Theorem~\ref{thm:general} gives the bound; it applies to earliest-start list scheduling by Lemma~\ref{lem:es-list}.
For $\rho=0.26$, $a_\rho=2/1.26=100/63$ and $2/(2-\rho)=2/1.74=100/87$. Then $a_\rho^2+4=25876/3969$ and $\sqrt{25876}=2\sqrt{6469}$, so $h(100/63)=\big(2-\frac{100}{63}+\frac{2\sqrt{6469}}{63}\big)/2=(13+\sqrt{6469})/63$ and $\frac{100}{87}h(100/63)=100(13+\sqrt{6469})/5481$ because $87\cdot63=5481$. By Lemma~\ref{lem:beta}, $\beta(100/63)=2\big/\big(\frac{226+2\sqrt{6469}}{63}\big)=63/(113+\sqrt{6469})$. The value of the infimum is obtained numerically.
\end{proof}

The bound of Theorem~\ref{thm:cc13} coincides with the closed form of~\cite{Jansen11} for $\rho=0.26$, and its infimum coincides with their asymptotic value $3.291913$. This is no coincidence: by Section~\ref{sec:cc13-alg} the algorithm is the one of~\cite{Jansen11}, the per-task factors of Lemmas~\ref{lem:cc13-time} and~\ref{lem:cc13-work} are those of~\cite[Lemma~4.2]{Jansen11}, and Theorem~\ref{thm:general} reproduces their min--max analysis for every $m$. The assumptions are also the same by Proposition~\ref{prop:concave}. Hence, with the parameters of~\cite{Jansen11}, the algorithm of~\cite{Chen13} matches, but does not improve, the ratio of~\cite{Jansen11}. For the same instances, speedup-aware rounding achieves $\approx2.973912$ (Theorem~\ref{thm:strict}). The parameters $\rho$ and $\mu$ of~\cite[Table~3]{Chen13} were chosen by the min--max analysis that Section~\ref{sec:cc13-fail} shows to be invalid; once the analysis is corrected they have no special role, and Theorem~\ref{thm:cc13} chooses the parameters by the corrected analysis.

\subsection{Remarks on the general case of Chen and Chu}
The ratio $2+\sqrt2$ of~\cite{Chen13} under \Aone, \Atwo\ and \Athree\ is not affected: it corresponds to $\rho=0$, that is, to rounding up only, and Theorem~\ref{thm:general} with $a=2$ and $b=1$ reproduces it for every $m$ (Remark~\ref{rem:old-rules}). One step of its proof needs a different justification. In the proof of~\cite[Lemma~4.4]{Chen13}, the case of a capped task with $l_j^*=w_j(x_j^*)/x_j^*\ge\mu$ uses the inequality $W_j(\lfloor l_j^*\rfloor)\le w_j(x_j^*)$. This inequality can fail when the processing times have a plateau: for $p=(1,1,1,0.75+\varepsilon,\dots)$ and $x^*=0.76$, we have $w(x^*)\approx2.92$ and $l^*\approx3.84$, but $W(3)=3$. The conclusion needed there, $p_j(\mu)\le(m/\mu)x_j^*$, remains true: the threshold rule satisfies $W(l')\le m\,x^*$ in all cases (when rounding up, $W(s)\le w(x^*)\le m\,x^*$ by Facts~\ref{fact:monotone} and~\ref{fact:mx}; when rounding down, $W(q+1)\le m\,p(q+1)<m\,x^*$), so Lemma~\ref{lem:capped-general} applies.

\FloatBarrier
\section{Common Speedup Models}\label{sec:models}
We check that five speedup models satisfy the conditions of Theorem~\ref{thm:strict} after removing useless allotments: the roofline, communication, Amdahl and general models, which are used in the literature on moldable task graphs~\cite{Benoit22,Perotin24}, and the power-law model, which serves as the typical example in~\cite{Jansen11} and is also considered in~\cite{Makarychev14}. Hence the ratio $\approx2.973912$ applies to them. Table~\ref{tab:models} lists the models.

\begin{table}[t]
\caption{Speedup models; $w>0$ is the sequential work of the task.}
\label{tab:models}
\vspace{2mm}
\centering
\begin{tabular}{lll}
\toprule
Model & $p(l)$ & Justification\\
\midrule
Roofline & $w/\min(l,\bar P)$, $\bar P\ge1$ an integer & Proposition~\ref{prop:three-models}\\
Amdahl & $w\,(\gamma+(1-\gamma)/l)$, $0\le\gamma<1$ & Proposition~\ref{prop:three-models}\\
Power law & $w\,l^{-d}$, $0<d\le1$ & Proposition~\ref{prop:three-models}\\
Communication & $w/l+c\,(l-1)$, $c\ge0$ & Proposition~\ref{prop:general-model}\\
General & $w\big((1-\gamma)/\min(l,\bar P)+\gamma\big)+c\,(l-1)$ & Proposition~\ref{prop:general-model}\\
\bottomrule
\end{tabular}
\end{table}

\begin{lem}[Removing useless allotments]\label{lem:useless}
Let $l^\circ$ be the smallest allotment at which the processing time of a task attains its minimum. Restricting the allotments of this task to $\{1,\dots,l^\circ\}$ does not change $\OPT$.
\end{lem}
\begin{proof}
Take any schedule in which the task uses $l>l^\circ$ processors, and let it use $l^\circ$ processors with the same start time instead. Its processing time $p(l^\circ)\le p(l)$ does not increase, so it occupies fewer processors for a shorter time and the capacity constraint still holds; it completes no later than before, so the precedence constraints still hold; and the makespan does not increase. Hence the restricted optimum is at most the original one, and it cannot be smaller because the restriction only removes options.
\end{proof}

The algorithm is applied to the restricted instance: in~\eqref{lp:range} the lower bound $p_j(m)$ becomes $p_j(l^\circ_j)$, and only the allotments $1,\dots,l^\circ_j$ are considered. The largest allotment of the task is then $l^\circ_j\le m$; the analysis uses only that allotments do not exceed $m$ (Fact~\ref{fact:mx} and (R3)), so it is not affected. The same restriction makes the LP of~\cite{Jansen11} well defined for concave speedups with a final plateau (Remark~\ref{rem:jz12}).

\begin{prop}[Three concave models]\label{prop:three-models}
The roofline, Amdahl and power-law models satisfy the conditions of Proposition~\ref{prop:concave}.
\end{prop}
\begin{proof}
The speedups $S(l)=p(1)/p(l)$ are $\min(l,\bar P)$, $1/(\gamma+(1-\gamma)/l)=l/(\gamma l+1-\gamma)$ (because $p(1)=w$) and $l^d$, respectively. Each is the restriction to the integers of a function $f$ on $[0,\infty)$ with $f(0)=0$ that is non-decreasing and concave; the chord slopes of a concave function between consecutive integers are non-increasing, $f(l)-f(l-1)\ge f(l+1)-f(l)$, so $S$ is concave on $\{0,\dots,m\}$. The function $\min(x,\bar P)$ is a minimum of two affine functions and hence concave. For $f(x)=x/(\gamma x+1-\gamma)$, $f'=(1-\gamma)/(\gamma x+1-\gamma)^2>0$ and $f''=-2\gamma(1-\gamma)/(\gamma x+1-\gamma)^3\le0$. For $f(x)=x^d$, $f'=d\,x^{d-1}>0$ for $x>0$ and $f''=d(d-1)x^{d-2}\le0$.
\end{proof}

\begin{prop}[Communication and general models]\label{prop:general-model}
Consider the general model
\[
p(l)=w\Big(\frac{1-\gamma}{\min(l,\bar P)}+\gamma\Big)+c\,(l-1),\qquad w>0,\ 0\le\gamma<1,\ \bar P\ge1\text{ an integer},\ c\ge0.
\]
Restricted to $\{1,\dots,l^\circ\}$ as in Lemma~\ref{lem:useless}, it satisfies \Aonec, \Atwo\ and \Athree. The communication model $p(l)=w/l+c(l-1)$ is the special case $\gamma=0$, $\bar P\ge m$.
\end{prop}
\begin{proof}
Write $d_l=p(l)-p(l+1)$ and $e_l=W(l+1)-W(l)$.
(1) For $l+1\le\bar P$, $p(l)=w(1-\gamma)/l+w\gamma+c(l-1)$, so $d_l=w(1-\gamma)\big(\frac1l-\frac1{l+1}\big)-c=\frac{w(1-\gamma)}{l(l+1)}-c$, which is strictly decreasing in $l$ because $w(1-\gamma)>0$. Moreover $W(l)=l\,p(l)=w(1-\gamma)+w\gamma\,l+c\,l(l-1)$, so $e_l=w\gamma+2cl\ge0$, which is non-decreasing in $l$.
(2) For $l\ge\bar P$, $p(l+1)-p(l)=c\ge0$, so $p$ is non-decreasing from $\bar P$ on.
(3) By (2), $l^\circ\le\bar P$. Let $l<l^\circ$. If $d_l\le0$, then by the monotonicity in (1), $d_k\le0$ for $l\le k\le\bar P-1$, and with (2) $p$ is non-decreasing from $l$ on; then the minimum of $p$ is attained at an allotment $\le l<l^\circ$, contradicting the choice of $l^\circ$. Hence $d_l>0$ for $l<l^\circ$, and $p$ strictly decreases on $\{1,\dots,l^\circ\}$; after the restriction this is \Aonec.
(4) \Aone\ follows from (3), and \Atwo\ from $e_l\ge0$. For \Athree, the marginal work between consecutive allotments $l<l^\circ$ is $\sigma_l=e_l/d_l$. Since $e_{l+1}\ge e_l\ge0$ and $0<d_{l+1}<d_l$, we get $\sigma_{l+1}=e_{l+1}/d_{l+1}\ge e_l/d_{l+1}\ge e_l/d_l=\sigma_l$.
\end{proof}

Only general instances satisfying \Aone, \Atwo\ and \Athree\ with plateaus in the middle, such as the task $p=(1,1,2/3)$ of Example~\ref{ex:tight}, need the ratio $3.016208$ of Theorem~\ref{thm:opt-a}.

\section{Limits of the Approach}\label{sec:limits}
Speedup-aware rounding gains from convexity. Without \Athree, no per-task rounding beats the trade-off of~\cite{Jansen06}; and in the framework of Section~\ref{sec:general}, even a perfect rounding cannot lead below $(3+\sqrt5)/2$.

\begin{prop}[No improvement under \Aone\ and \Atwo\ alone]\label{prop:a12}
Without \Athree, the LP stays a relaxation only if the work function of each task is replaced by the lower convex hull $\underline w$ of the points $\{(p(l),W(l))\}$ (Remark~\ref{rem:hull}). For every $a>1$ and $\eta>0$ there are a task satisfying \Aonep\ and \Atwo\ and a value $x^*\in[p(m),p(1)]$ such that every allotment with $p(l)\le a\,x^*$ has $W(l)\ge\big(a/(a-1)-\eta\big)\,\underline w(x^*)$. Hence every per-task rounding with time factor $a$ has work factor at least $a/(a-1)$, the trade-off of~\cite{Jansen06}, and the analysis of Section~\ref{sec:general} cannot improve the ratio $\approx4.7306$ in this setting.
\end{prop}
\begin{proof}
Choose an integer $k\ge2$, let $\varepsilon=1/(4k^2)$, choose an integer $m\ge\max\{k^2,2ak\}$, let $W_k=k(1-\varepsilon(k-1))$, and define
\[
p(l)=1-\varepsilon(l-1)\ \ (1\le l\le k),\qquad p(l)=W_k/l\ \ (k<l\le m).
\]
The task satisfies \Aonep: for $l\le k$ the processing times strictly decrease, $p(k+1)=W_k/(k+1)=k\,p(k)/(k+1)<p(k)$, and $W_k/l$ strictly decreases for $l>k$. It also satisfies \Atwo: for $l<k$, $W(l+1)-W(l)=(l+1)(1-\varepsilon l)-l(1-\varepsilon(l-1))=1-2\varepsilon l\ge1-1/(2k)>0$; for $l\ge k$, $W(l)=W_k$.

Next we show that the lower convex hull is a single segment. Let $\ell(x)=1+(W_k-1)(1-x)/(1-p(m))$ be the line through $(p(1),W(1))=(1,1)$ and $(p(m),W_k)$. For $l>k$, $\ell(p(l))\le W_k$ is equivalent to $(1-p(l))/(1-p(m))\le1$, which holds since $p(l)\ge p(m)$ (here $W_k>1$). For $2\le l\le k$, we have $1-p(l)=\varepsilon(l-1)\le\varepsilon k$, $1-p(m)\ge1/2$ (as $p(m)\le k/m\le1/2$) and $W_k-1<k$, so $\ell(p(l))\le1+2\varepsilon k^2=3/2$, whereas $W(l)\ge W(2)=2(1-\varepsilon)>3/2$. All points lie on or above $\ell$, so $\underline w=\ell$ on $[p(m),1]$.

Now let $x^*=p(k)/a$. Then $x^*<1$, and $x^*\ge(1-1/(4k))/a\ge7/(8a)>1/(2a)\ge k/m\ge p(m)$.
Every allotment with $p(l)\le a\,x^*=p(k)$ has $l\ge k$, since $p$ is strictly decreasing, and hence $W(l)\ge W_k$ by \Atwo. Its work factor is therefore at least $W_k/\underline w(x^*)=W_k/\ell(x^*)=1\big/\big(1/W_k+(1-1/W_k)(1-x^*)/(1-p(m))\big)$. As $k\to\infty$, $W_k\to\infty$, $x^*\to1/a$ and $p(m)\le k/m\le1/k\to0$, so this bound tends to $1/(1-1/a)=a/(a-1)$. Choosing $k$ large enough proves the claim.
\end{proof}

Proposition~\ref{prop:a12} concerns a single relaxation in which the work of each task is bounded from below by the lower convex hull of its own work function, combined with a per-task rounding. Within the framework of Section~\ref{sec:general}, progress under \Aone\ and \Atwo\ therefore requires a relaxation with stronger work bounds. The iterative method of~\cite{Chen18} solves a sequence of LPs whose work bounds are taken from the previous LP solution, so Proposition~\ref{prop:a12} does not apply to it directly; we do not re-examine its analysis here, and Table~\ref{tab:summary} lists its ratio as reported.

\begin{prop}[Limit of the framework]\label{prop:frame}
For every $a>1$ and $b\ge1$, the bound $a+b\,h(a)$ of Theorem~\ref{thm:general} exceeds $(3+\sqrt5)/2\approx2.618$.
\end{prop}
\begin{proof}
$a+b\,h(a)\ge a+h(a)=1/\beta(a)$ by Lemma~\ref{lem:beta}(c), and $1/\beta$ is strictly increasing by Lemma~\ref{lem:deriv}(b), so $a+b\,h(a)>1/\beta(1)=(1+2+\sqrt5)/2=(3+\sqrt5)/2$.
\end{proof}

The bound of Theorem~\ref{thm:general} cannot be far from the truth for the algorithm itself, because the cap alone is costly.

\begin{prop}[Lower bounds]\label{prop:lower}
\textup{(a)} The integrality gap of $(\mathrm{P})$ is at least $2$, already for tasks without speedup.
\textup{(b)} On a single task with $p(l)=m/l$ for $l=1,\dots,m$, every algorithm that caps the allotments at $\mu$ has makespan at least $m/\mu$, while $C^*=\OPT=1$. In particular, for every $a>1$ the approximation ratio of $\SAR(a,\lceil\beta(a)m\rceil)$ over all $m$ is at least $1/\beta(a)=a+h(a)$; this equals $(31+\sqrt{493})/18\approx2.955756$ for $a=13/9$ and $\approx2.830972$ for $a=a_s$.
\textup{(c)} The ratio $2+\sqrt2$ of rounding up only ($a=2$, $b=1$, the general rule of~\cite{Chen13}) is tight.
\end{prop}
\begin{proof}
(a) Let $m\ge k\ge1$, and let every task satisfy $p_j(1)=\dots=p_j(m)=1$; such tasks satisfy \Aone, \Atwo\ and \Athree\ with $K_j=1$. Take a chain $J_1\prec\dots\prec J_k$ and $mk$ further tasks, each a successor of $J_k$. In $(\mathrm{P})$ every $x_j$ equals $1$ and $w_j\ge1$. The longest path has $k+1$ tasks, so $C^*\ge k+1$, and choosing $C_j$ as the number of tasks on a longest path ending in $j$, $w_j=1$ and $L=C=k+1$ is feasible because the total work is $k+mk\le m(k+1)$; hence $C^*=k+1$. In every schedule the chain needs time $k$, and the $mk$ successors start after it and have total work at least $mk$, so they need time at least $k$; running the chain and then the successors on one processor each attains $\OPT=2k$. The ratio $2k/(k+1)$ tends to $2$. For tasks without speedup, $(\mathrm{P})$ is the classical lower bound $\max\{\text{longest path},\ \text{total work}/m\}$ for $P|prec|C_{\max}$, whose integrality gap is well known to approach $2$; we include the example for completeness. It shows that no analysis against $C^*$ can prove a ratio below $2$.

(b) The task has $W(l)=m$ for every $l$, so it satisfies \Aonep, \Atwo\ and \Athree, and its speedup $S(l)=l$ is concave. In $(\mathrm{P})$, $C\ge C_1\ge x_1\ge p(m)=1$, and $x_1=1$, $w_1=m$, $C_1=L=C=1$ is feasible, so $C^*=1$; running the task on $m$ processors gives $\OPT=1$. Every allotment $l\le\mu$ has $p(l)=m/l\ge m/\mu$. For $\mu=\lceil\beta(a)m\rceil<\beta(a)m+1$ the ratio $m/\mu$ exceeds $m/(\beta(a)m+1)$, which tends to $1/\beta(a)$ as $m\to\infty$; Lemma~\ref{lem:beta}(c) gives $1/\beta(a)=a+h(a)$, and $1/\beta(13/9)=(31+\sqrt{493})/18$. More generally, every cap with $\mu\le c\,m$ gives the ratio at least $1/c$ on this instance. The cap chosen as in Remark~\ref{rem:mu} satisfies $\mu/m\to\beta(a)$, because by Remark~\ref{rem:why-beta} the limit of $\max\{A,B\}$ has its unique minimum at $\beta(a)$; hence the lower bound $1/\beta(a)$ holds for that cap as well. (For $a=13/9$ and $m=1024$, for instance, the program for Table~\ref{tab:finite} gives $\mu=347=\lceil\beta(a)m\rceil$.)

(c) By (b) with $a=2$, the ratio is at least $1/\beta(2)=2+\sqrt2$, which is the upper bound of Theorem~\ref{thm:general} for $a=2$ and $b=1$. The same holds for the cap $\hat\mu^*=(2m-\sqrt{2m^2-2m})/2$ of~\cite{Chen13}, since $\hat\mu^*/m\to1-1/\sqrt2=\beta(2)$.
\end{proof}

For the ratios $3.016208$ and $2.973912$, the price of rounding $(b-1)h$ accounts for only about $0.060$ and $0.143$, respectively; the rest is $1/\beta(a)$, which by Proposition~\ref{prop:lower}(b) is a lower bound for the algorithm itself. The approximation ratio of $\SAR(a,\lceil\beta(a)m\rceil)$ therefore lies in $[2.955756,\,3.016208]$ for $a=13/9$ and in $[2.830972,\,2.973912]$ for $a=a_s$.

\begin{rem}[Directions for improvement]\label{rem:directions}
(1) The inequality \Win\ counts a single busy processor during $T_1$. A better lower bound on the load during $T_1$, for instance from the work of the tasks on the heavy path, would allow a larger cap $\mu$ and thereby decrease the term $1/\beta(a)$; by Proposition~\ref{prop:lower}(b), this term cannot be reduced without changing the cap. This seems the most promising direction. (2) Proposition~\ref{prop:optimal} shows that the per-task rounding is already optimal, so further progress must come from a finer analysis of list scheduling or from a stronger LP with inequalities that couple several tasks. (3) The min--max program~(17) of~\cite{Jansen11} coincides with Lemma~\ref{lem:inner} for $a=2/(1+\rho)$ and $b=2/(2-\rho)$, so it offers no further gain. A finer analysis could let the time and work factors depend on the type of each task, for instance by treating blocks with $r\ge\alpha$, which are always rounded up at no extra work, separately in \Hin\ and \Win; we have not pursued this. (4) The integrality gap of $(\mathrm{P})$ is at least $2$ (Proposition~\ref{prop:lower}(a)) and at most $3.016208$ under \Aone--\Athree\ (Theorem~\ref{thm:opt-a}); its exact value is not known. The ratio of $\SAR(a,\lceil\beta(a)m\rceil)$ lies between $1/\beta(a)$ and $a+b(a)h(a)$; instances on which the losses of rounding and of capping add up would close this gap.
\end{rem}

\section{Numerical Checks}\label{sec:numerics}
All results above have analytical proofs. The proofs use numbers only in the form of rational bounds on square roots, such as $\sqrt{493}$ in Theorem~\ref{thm:opt-a}; the minimizers $a_s$ and $\rho\approx0.26192$ are reported numerically, but they are not needed for the closed-form bounds at $a=9/7$ and $\rho=0.26$. As an independent check we implemented the formulas and algorithms in Python (with NumPy and SciPy/HiGHS); the programs are available as ancillary files. All checks passed:
\begin{itemize}
\item the formula $1+F(q,s;r)$ agrees with the directly computed work ratio of Lemma~\ref{lem:rounding} on 200{,}000 random blocks (maximum deviation $4.4\cdot10^{-16}$), and Lemmas~\ref{lem:F-s} and~\ref{lem:singleton} hold on 200{,}000 random parameter sets each;
\item the closed forms of Theorems~\ref{thm:closed} and~\ref{thm:strict} agree with a brute-force supremum over all $q\le200$; Proposition~\ref{prop:21} and Lemma~\ref{lem:crossing} are confirmed; and a grid search of $R(a)$ over $(1,3]$ finds no value below $R(13/9)$;
\item the inner LP of Lemma~\ref{lem:inner}, solved exactly on 3{,}000 random parameter sets, never exceeds $\max\{A,B\}$; for $m=2,\dots,20{,}000$ and $\mu=\lceil\beta(a)m\rceil$ the largest values of $\max\{A(\mu),B(\mu)\}$ are $3.044153$, $3.016203$ and $2.973901$ for the parameters of Corollary~\ref{coro:three-half}, Theorem~\ref{thm:opt-a} and Proposition~\ref{prop:strict-opt}, all below the corresponding bounds;
\item Lemmas~\ref{lem:no-idle} and~\ref{lem:es-list} hold in 12{,}727 checks on 3{,}000 random schedules;
\item on 800 random instances (LP, rounding, cap, list scheduling) the largest ratio $\Cmax/C^*$ is $1.8843$, and on small instances with exact optimum the largest ratio $\Cmax/\OPT$ is $1.7518$;
\item for Section~\ref{sec:cc13}, on 200{,}000 random tasks satisfying \Aonec, \Atwo\ and \Athree, $72{,}201$ of them with a final plateau, the threshold rule chooses only non-dominated allotments and never exceeds the factors $2/(1+\rho)$ and $2/(2-\rho)$ of Lemmas~\ref{lem:cc13-time} and~\ref{lem:cc13-work}; the program of~\cite[Lemma~4.15]{Chen13} with $x_p\le m$ has the value $3.5523$ for $m=8$, $\mu=4$, $\rho=0.5921$; and the guarantee of Theorem~\ref{thm:cc13} holds for all $m\le2\cdot10^5$;
\item for Proposition~\ref{prop:concave}, on 200{,}000 random tasks satisfying \Aonec, conditions \Atwo\ and \Athree\ hold exactly when the speedup is concave;
\item the examples of Proposition~\ref{prop:a12} approach $a/(a-1)$, and the models of Section~\ref{sec:models} satisfy \Aonec, \Atwo\ and \Athree\ for 20{,}000 random parameter sets each;
\item for Proposition~\ref{prop:lower}, the implementation (LP, rounding, cap, list scheduling) gives $C^*=k+1$ and $\Cmax=\OPT=2k$ on the instances of part~(a) for $k=m\le6$, and $\Cmax/\OPT=m/\lceil\beta(a)m\rceil$ on the instance of part~(b) for all $m\le64$, with maxima $2.952381$ for $a=13/9$ and $2.823529$ for $a=a_s$; over $m\le2\cdot10^5$, $\max_m m/\lceil\beta(a)m\rceil$ agrees with $1/\beta(a)$ to six decimals.
\end{itemize}

Table~\ref{tab:finite} lists the guarantees $\min_\mu\max\{A(\mu),B(\mu)\}$ of Lemma~\ref{lem:inner} for small $m$ (Remark~\ref{rem:mu}). For a given $m$ one may also choose $a$; rows (I$^*$) and (II$^*$) do so over the ranges where $b(a)$ is known in closed form, $a\in[13/9,2]$ and $a\in(1,2]$, with $a$ rounded to four decimals (the ancillary program lists the chosen $a$ and $\mu$). Under \Aone--\Athree, speedup-aware rounding with the best $a$ equals rounding up only for $m=2$ and is better for every larger $m$ in the table. Under \Aonec\ (and hence under \Aonep), it is better than the threshold rule of~\cite{Jansen11} for every $m$ in the table.

\begin{table}[t]
\caption{Guarantees $\min_{1\le\mu\le(m+1)/2}\max\{A(\mu),B(\mu)\}$ on $\Cmax/C^*$, grouped by assumptions. Under \Aone--\Athree: (I) speedup-aware rounding with $a=13/9$; (I$^*$) speedup-aware rounding with the best $a\in[13/9,2]$; (III) rounding up only ($a=2$, $b=1$), the general rule of~\cite{Chen13}. Under \Aonec, \Atwo, \Athree: (II) speedup-aware rounding with $a=a_s$; (II$^*$) speedup-aware rounding with the best $a\in(1,2]$; (IV) the threshold rule with $\rho=0.26$ under \Aonep, that is, the algorithm of~\cite{Jansen11} (Theorem~\ref{thm:cc13}).}
\label{tab:finite}
\vspace{2mm}
\centering
\small
\setlength{\tabcolsep}{4.5pt}
\begin{tabular}{lccccccccc}
\toprule
Rule & $m=2$ & $m=4$ & $m=8$ & $m=16$ & $m=32$ & $m=64$ & $m=128$ & $m=1024$ & $m\to\infty$\\
\midrule
(I) & 2.0400 & 2.3496 & 2.7200 & 2.8259 & 2.9604 & 2.9686 & 2.9936 & 3.0131 & 3.0162\\
(I$^*$) & 2.0000 & 2.3496 & 2.6882 & 2.8259 & 2.9465 & 2.9655 & 2.9936 & 3.0131 & 3.0162\\
(III) & 2.0000 & 2.6667 & 3.0000 & 3.2000 & 3.3043 & 3.3684 & 3.3846 & 3.4116 & 3.4142\\
\midrule
(II) & 2.0924 & 2.3131 & 2.7758 & 2.8011 & 2.8883 & 2.9235 & 2.9541 & 2.9715 & 2.9739\\
(II$^*$) & 2.0000 & 2.3131 & 2.6882 & 2.7711 & 2.8834 & 2.9201 & 2.9490 & 2.9705 & 2.9739\\
(IV) & 2.1494 & 2.5908 & 2.8659 & 3.1149 & 3.1870 & 3.2612 & 3.2675 & 3.2883 & 3.2919\\
\bottomrule
\end{tabular}
\end{table}

Finally, we compared four rules on the same random instances (Table~\ref{tab:compare}): speedup-aware rounding with $a=13/9$ and with $a=a_s$, the threshold rule with $\rho=0.26$, and rounding up only. For each $m\in\{8,32\}$ we generated 60 random DAGs with $n=30$ tasks; each instance contains an arc $(i,j)$, $i<j$, independently with probability $0.05$, $0.1$ or $0.2$, chosen at random per instance. In the first class every task follows one of the five models of Section~\ref{sec:models}, chosen at random, with $w\in[1,100]$, $\gamma\in[0,0.3]$, $d\in[0.3,1]$, $c\in[0,0.5]$ and $\bar P\in\{1,\dots,m\}$; the allotments beyond the first minimizer receive the time of that minimizer, which is equivalent to the restriction of Lemma~\ref{lem:useless}. In the second class every task satisfies \Aone--\Athree\ with plateaus inside the range of allotments; it is built block by block, the next non-dominated allotment being $s+1$ with probability $0.6$ and $s+\delta$ with $\delta\in\{2,3,4\}$ otherwise, and its processing time is chosen so that Fact~\ref{fact:ratio} holds and the marginal works do not decrease. The random seed is fixed in \texttt{compare\_rules.py}. All rules use the same LP solution and their own cap $\mu=\lceil\beta(a)m\rceil$. The random instances are far from the worst case, since every ratio $\Cmax/C^*$ is below $1.93$, and the rules behave similarly: the means differ by at most $0.08$. $\SAR(a_s)$ has the smallest mean in three of the four classes and rounding up only in the remaining one. Instance by instance, for the model tasks with $m=32$, $\SAR(a_s)$ gives a smaller makespan than the threshold rule on 38 instances and a larger one on 19. What distinguishes the rules is their worst-case guarantee rather than their typical behavior.

\begin{table}[htbp]
\caption{Mean and maximum (in parentheses) of $\Cmax/C^*$ over 60 random instances with $n=30$ tasks for each class and each $m$; all rules use the same LP solution and the cap $\mu=\lceil\beta(a)m\rceil$ (ancillary program \texttt{compare\_rules.py}).}
\label{tab:compare}
\vspace{2mm}
\centering
\small
\begin{tabular}{llcccc}
\toprule
tasks & $m$ & $\SAR(13/9)$ & $\SAR(a_s)$ & threshold $\rho=0.26$ & rounding up only\\
\midrule
models of Section~\ref{sec:models} & 8 & 1.306 (1.599) & 1.301 (1.548) & 1.308 (1.599) & 1.326 (1.544)\\
models of Section~\ref{sec:models} & 32 & 1.287 (1.781) & 1.276 (1.778) & 1.301 (1.906) & 1.354 (1.924)\\
\Aone--\Athree\ with plateaus & 8 & 1.251 (1.660) & 1.255 (1.660) & 1.258 (1.660) & 1.249 (1.660)\\
\Aone--\Athree\ with plateaus & 32 & 1.071 (1.398) & 1.060 (1.512) & 1.067 (1.398) & 1.095 (1.392)\\
\bottomrule
\end{tabular}
\end{table}

\FloatBarrier
\section{Concluding Remarks}\label{sec:conclusion}
We proposed speedup-aware rounding for scheduling malleable tasks with precedence constraints. The rule rounds a task to its slower allotment exactly when this costs at most a factor $a$ in time, and we computed its worst-case work increase $b(a)$ exactly and showed that no per-task rule does better. A short duality argument turns any per-task guarantee $(a,b)$ into the approximation ratio $a+b\,h(a)$ for every number of processors. Under the monotonous penalty assumptions and convex work this gives $13(30+\sqrt{493})/225\approx3.0162$, improving $2+\sqrt2$; since a single task with linear speedup forces the ratio $1/\beta(13/9)\approx2.9558$, the analysis is within $0.061$ of tight. For concave speedup functions, equivalently when processing times strictly decrease until they reach their minimum, a class that contains the common roofline, communication, Amdahl and power-law models, it gives $\approx2.9739$, improving the ratio $\approx3.2919$ of~\cite{Jansen11}. With the same tools we corrected the analysis of the strictly decreasing case of~\cite{Chen13}: there its algorithm coincides with that of~\cite{Jansen11} up to the parameters, and with parameters chosen by the corrected analysis it achieves $\approx3.2919$, the ratio of~\cite{Jansen11}, also for concave speedups with a final plateau, whose plateau allotments can be removed. Without convexity, per-task rounding cannot improve the known trade-off, and within the framework no rounding leads below $(3+\sqrt5)/2$. Closing the gap to the lower bounds requires a sharper analysis of list scheduling together with a different cap, or a stronger relaxation.

\section*{Acknowledgment}
The author used Claude (Anthropic), a large language model, as an assistant in this work: to expand and check the proofs, to write the numerical verification programs, to survey recent literature, and to draft and edit the manuscript. The author reviewed and edited all AI-assisted content and takes full responsibility for the content of this paper.


\end{document}